\documentclass[aps, twocolumn, prl, amsmath, superscriptaddress]{revtex4-1}
\usepackage{graphicx}
\usepackage{dcolumn}
\usepackage{bm}
\usepackage{amsmath}
\usepackage{subfigure}
\usepackage{amsfonts}
\usepackage{xcolor}
\usepackage{appendix}
\usepackage{multirow}
\usepackage[colorlinks,
            linkcolor=blue,
            anchorcolor=red,
            citecolor=blue
            ]{hyperref}
\usepackage{amsthm}
\usepackage{setspace}
\usepackage{lineno}

\usepackage[draft]{changes} 

\begin{document}

\title{Path-Independent Quantum Gates with Noisy Ancilla}
\author{Wen-Long Ma}
\affiliation{Pritzker School of Molecular Engineering, University of Chicago, Illinois 60637, USA}
\affiliation{Department of Applied Physics and Physics, Yale University, New Haven, Connecticut 06511, USA}
\affiliation{Yale Quantum Institute, Yale University, New Haven, Connecticut 06511, USA}
\author{Mengzhen Zhang}
\affiliation{Pritzker School of Molecular Engineering, University of Chicago, Illinois 60637, USA}
\affiliation{Department of Applied Physics and Physics, Yale University, New Haven, Connecticut 06511, USA}
\affiliation{Yale Quantum Institute, Yale University, New Haven, Connecticut 06511, USA}
\author{Yat Wong}
\affiliation{Pritzker School of Molecular Engineering, University of Chicago, Illinois 60637, USA}
\author{Kyungjoo Noh}
\affiliation{Pritzker School of Molecular Engineering, University of Chicago, Illinois 60637, USA}
\affiliation{Department of Applied Physics and Physics, Yale University, New Haven, Connecticut 06511, USA}
\affiliation{Yale Quantum Institute, Yale University, New Haven, Connecticut 06511, USA}
\author{Serge~Rosenblum}
\affiliation{Department of Applied Physics and Physics, Yale University, New Haven, Connecticut 06511, USA}
\affiliation{Yale Quantum Institute, Yale University, New Haven, Connecticut 06511, USA}
\affiliation{Department of Condensed Matter Physics, Weizmann Institute of Science, Rehovot, Israel}
\author{Philip Reinhold}
\affiliation{Department of Applied Physics and Physics, Yale University, New Haven, Connecticut 06511, USA}
\affiliation{Yale Quantum Institute, Yale University, New Haven, Connecticut 06511, USA}
\author{Robert J. Schoelkopf}
\affiliation{Department of Applied Physics and Physics, Yale University, New Haven, Connecticut 06511, USA}
\affiliation{Yale Quantum Institute, Yale University, New Haven, Connecticut 06511, USA}
\author{Liang Jiang}
\affiliation{Pritzker School of Molecular Engineering, University of Chicago, Illinois 60637, USA}
\affiliation{Department of Applied Physics and Physics, Yale University, New Haven, Connecticut 06511, USA}
\affiliation{Yale Quantum Institute, Yale University, New Haven, Connecticut 06511, USA}

\date{\today }

\begin{abstract}
Ancilla systems are often indispensable to universal control of a nearly isolated quantum system. However, ancilla systems are typically more vulnerable to environmental noise, which limits the performance of such ancilla-assisted quantum control. To address this challenge of ancilla-induced decoherence, we propose a general framework that integrates quantum control and quantum error correction, so that we can achieve robust quantum gates resilient to ancilla noise. We introduce the path independence criterion for fault-tolerant quantum gates against ancilla errors. As an example, a path-independent gate is provided for superconducting circuits with a hardware-efficient design.
\end{abstract}

\maketitle


\maketitle


An outstanding challenge of quantum computing is building quantum devices with both excellent coherence and reliable universal control \cite{Nielsen2010,Glaser2015,Vandersypen2005}. For good coherence, we may choose physical systems with low dissipation (e.g., superconducting cavities \cite{Devoret2013,Schuster2007,Leghtas2013} and nuclear spins \cite{Kane1998,Dutt2007,Maurer2012,Liu2017}) or further boost the coherence with active quantum error correction \cite{Ofek2016,Waldherr2014}.
As we improve the coherence by better isolating the central system from the noisy environment, it becomes more difficult to process information stored in the central system. To control the nearly isolated central system, we often introduce an ancilla system (e.g., transmon qubits \cite{Koch2007,Krastanov2015,Heeres2015} and electron spins \cite{Dutt2007,Maurer2012}) that is relatively easy to control, but the ancilla system typically suffers more decoherence than the central system, limiting the fidelity of the ancilla-assisted quantum operations. Therefore, it is crucial to develop quantum control protocols that are fault-tolerant against ancilla errors.

For noise with temporal or spatial correlations, we can use techniques of dynamical decoupling \cite{Viola1999,Liu2013,Taminiau2014} or decoherence-free encoding \cite{Lidar1998,Bacon1999} to achieve noise-resilient control of the central system. When the noise has no correlations (e.g., Markovian noise), we need active quantum error correction (QEC) to extract the entropy. For qubit systems, a common strategy to suppress ancilla errors is to use the transversal approach \cite{Nielsen2010,Knill1998,Gottesman2010,DiVincenzo1996,Eastin2009,Fowler2012,Gottesman2016}, which may have a significant hardware overhead and cannot provide universal control \cite{Nielsen2010}, and it is desirable to have a hardware-efficient approach to fault-tolerant operations against ancilla errors \cite{Chao2018a,Chao2018b,Chamberland1,Chamberland2,Chamberland3,Chamberland4}.
Different from qubit systems, each bosonic mode has a large Hilbert space that can encode quantum information using various bosonic quantum codes as demonstrated in recent experiments \cite{Ofek2016,Fluhmann2019,Campagne-Ibarcq2019,Hu2019}. However, there is no simple way to divide the bosonic mode into separate subsystems, which prevents us from extending the transversal approach to the bosonic central system. Ancilla errors can propagate to the bosonic mode and compromise the encoded quantum information \cite{Heeres2017}. Nevertheless, a recent experiment with a hardware-efficient three-level ancilla demonstrated fault-tolerant readout of an error syndrome of the central system against the decay of the ancilla \cite{Rosenblum2018}. Moreover, the error-transparent gates for QEC codes (using control Hamiltonian commuting with errors) have been proposed \cite{Vy2013,Kapit2018, Ma2019} to achieve quantum operations insensitive to errors, but it is typically very demanding to fulfill the error-transparent condition while performing non-trivial quantum gates. Therefore, there is an urgent need of a general theoretical framework that integrates quantum control and quantum error correction, to guide the design of hardware-efficient robust quantum operations against ancilla errors.

\begin{figure}
\includegraphics[width=3.4in]{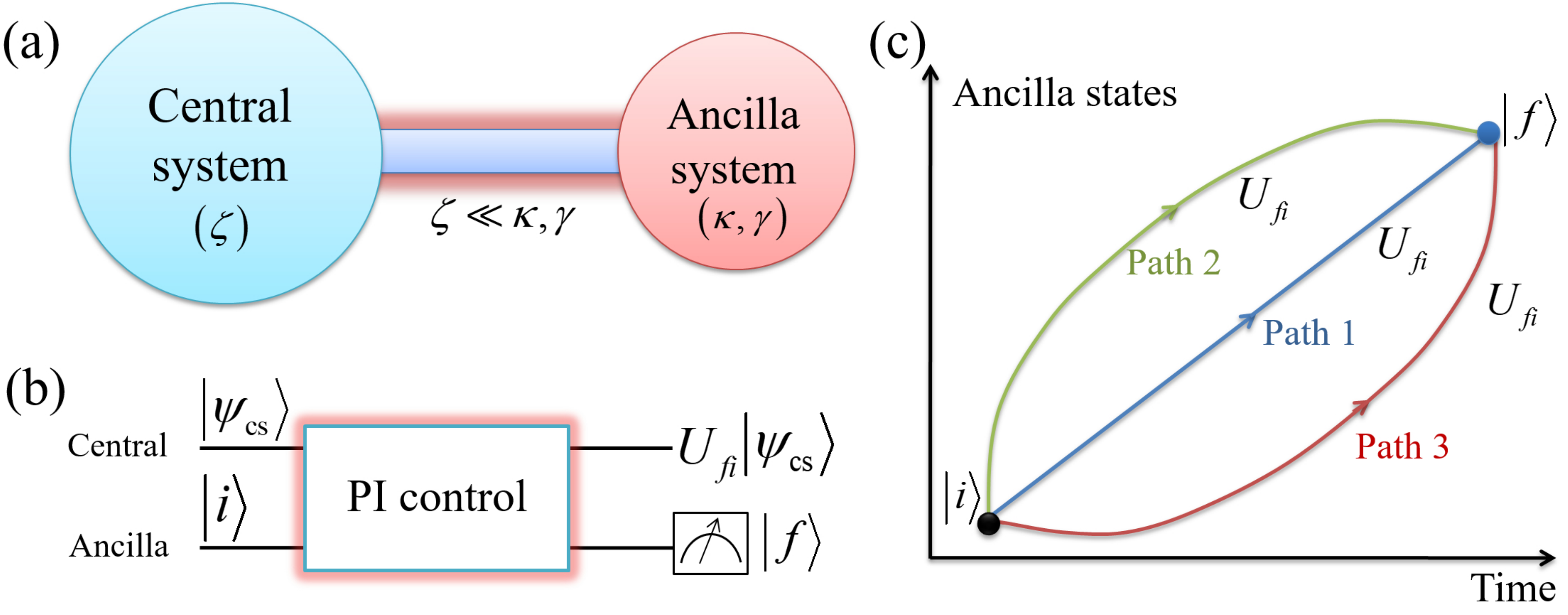}
\caption{(a) Schematic of a central system with good coherence coupled to an ancilla system with poor coherence. The ancilla dephasing rate $\kappa$ and relaxation rate $\gamma$ of are much larger than the decoherence rate $\zeta$ of the central system. (b),(c) For PI control, the central system undergoes a unitary evolution $U_{fi}$ for the ancilla starting from state $|i\rangle$ and finally measured in $|f\rangle$, regardless of the ancilla paths induced by the control and ancilla error events.}
\label{schematic}
\end{figure}

In this letter, we provide a general criterion for fault-tolerant quantum gates on the central system robust against ancilla errors [Fig. \ref{schematic}{\color{blue}(a)}]. Our general criterion of path-independence (PI) requires that for given initial and final ancilla states, the central system undergoes a unitary gate independent of the specific ancilla path induced by control drives and ancilla error events  [Fig. \ref{schematic}{\color{blue}(b)} and {\color{blue}(c)}]. For a subset of final ancilla states, the desired quantum gate on the central system is successfully implemented, while other final ancilla states herald a failure of the attempted operation,
but the central system still undergoes a deterministic unitary evolution without loss of coherence. Thus we may repeat our attempts of PI gates on the central system until the gate succeeds.
As an application of our general criterion, a PI design of the photon-number selective phase (SNAP) gates \cite{Krastanov2015,Heeres2015} is provided for universal control and quantum error correction of superconducting circuits. Moreover, the error-transparent gates \cite{Vy2013,Kapit2018, Ma2019} are also shown to be a special class of PI gates.


\textit{Ancilla-assisted quantum control}.--- Suppose we intend to implement some unitary gate on the central system assisted by a $d$-level ancilla system \cite{footnote3}. The total Hamiltonian is
\begin{eqnarray}
    H_{\rm tot}(t)=H_{0}+H_{\rm c}(t),
\end{eqnarray}
where $H_{\rm c}(t)$ is the control Hamiltonian and $H_{0}=H_{\rm as}+H_{\rm cs}+H_{\rm int}$ is the static Hamiltonian with contributions from the ancilla system, the central system and their interaction, correspondingly.
We assume that $[H_{\rm as},H_{\rm int}]=0$, so that $H_{\rm int}$ preserves the eigenbasis $\{|m\rangle\}_{m=0}^{d-1}$ of the ancilla ($H_{\rm as}|m\rangle=\varepsilon_{m}|m\rangle$).
The static Hamiltonian $H_0$ can be diagonalized in the eigenbasis of the ancilla,
\begin{eqnarray}\label{H0}
    H_{0}=\sum_{m=0}^{d-1} |m\rangle \langle m|\otimes (\varepsilon_m+H_{\rm cs}+H_{{\rm int},m}),
\end{eqnarray}
with $H_{{\rm int},m}=\langle m|H_{\rm int}|m\rangle$. The propagator for total system in the ancilla eigenbasis is
\begin{equation}
\begin{split}
    U(t_2,t_1)=&\mathcal{T}\mathrm{exp}\left(-i\int_{t_1}^{t_2} H_{\rm tot}(t')dt'\right) \\
    =&\sum_{m,n}\eta_{mn}(t_2,t_1)|m\rangle\langle n|\otimes V_{mn}(t_2,t_1),
\end{split}
\end{equation}
where $\mathcal{T}$ is the time-ordering operator, $\eta_{mn}(t_2,t_1)$ is a complex function and and $V_{mn}(t_2,t_1)$ is an operator on the central system. For pre-selection of the ancilla on state $|i\rangle$ at time $t_1$ and post-selection on $|f\rangle$ at $t_2$, the central system undergoes a quantum operation $V_{mn}(t_2,t_1)$.

\textit{Markovian ancilla noise}.--- We assume that the central system suffers much weaker noise than the ancilla and therefore can be regarded as noise-free within the ancilla coherence time [Fig. \ref{schematic}{\color{blue}(a)}]. Suppose the ancilla suffers from Markovian noise and the dynamics of the total system is
\begin{eqnarray}\label{Mar}
\frac{d{\rho}}{dt}=i[\rho, H_{\rm {tot}}(t)] +\left(\sum_l\mathcal{D}[\sqrt{\kappa_l}L_{l}] +\sum_j\mathcal{D}[\sqrt{\gamma_j}J_{j}]\right)\rho,  \nonumber \\
\end{eqnarray}
where $D[A]\rho=A\rho A^{\dagger}-\{A^{\dagger}A,\rho\}/2$ is the Lindbladian dissipator, $\{L_l\}$/$\{J_j\}$ are the Lindblad operators describing the ancilla dephasing/relaxation errors ($L_l=\sum_{m=0}^{d-1} \Delta_{l}^{(m)}|m\rangle\langle m|$ with $\Delta_{l}^{(m)}\in\mathbb{C}$, $J_j=|m_j\rangle\langle n_j|$ with $m_j, n_j\in[0,d-1]$ and $m_j\neq n_j$), and $\kappa_l/\gamma_j$ is the dephasing/relaxation rate. The ancilla dephasing and relaxation errors can be unified into a general class of ancilla errors \cite{SI}.

The Liouville superoperator $\mathcal{L}(t)$ generating the Markovian dynamics in Eq. (\ref{Mar}) can be divided into two parts \cite{Plenio1998},
\begin{eqnarray}
\frac{d{\rho}}{dt} =\mathcal{L}(t)\rho(t) = \left( {{\mathcal{L}_{\mathrm{eff}}}(t)
 + \mathcal{S}} \right)\rho(t),
\end{eqnarray}
where ${\mathcal{L}_{\mathrm{eff}}}\rho = i(\rho H_{\mathrm{eff}}^{\dagger}-H_{\mathrm{eff}}\rho)$ represents the no-jump evoultion with ${H_{\mathrm{eff}}}\left( t \right) = {H_{\rm tot}}\left( t \right) - \frac{i}{2}(\sum_l \kappa_l{L_l^\dag } {L_l}+\sum_j \gamma_j{J_j^\dag } {J_j})$, and $\mathcal{S}\rho = \sum_l\kappa_l{L_l} \rho L_l^\dag+\sum_j \gamma_j{J_j} \rho J_j^\dag$ represents the quantum jumps associated with the no-jump evolution. The propagator for the whole system can be represented by the generalized Dyson expansion as
\begin{eqnarray}\label{Dyson}
 \rho(t)=\sum_{p=0}^{\infty}\mathcal{G}_p(t,0)\rho (0),
\end{eqnarray}
with
\begin{eqnarray}
    \mathcal{G}_0(t,0)=\mathcal{W}(t,0),
\end{eqnarray}
\begin{equation}\label{Gp}
\begin{split}
 &\mathcal{G}_p(t,0)={\int_0^t {d{t_p}} } \cdots \int_0^{{t_3}} {d{t_{2}}} \int_0^{{t_2}} {d{t_1}}
  \mathcal{W}\left( {t,{t_p}} \right) \\
 &\times\mathcal{S}\cdots
 \mathcal{S}\mathcal{W}\left( {{t_2},{t_{1}}} \right)\mathcal{S} \mathcal{W}\left( {{t_1},0}\right), ~~p\geq1,
\end{split}
\end{equation}
where $\rho (0)=|m\rangle\langle m|\otimes \rho_{\rm cs}$ with $m\in[0,d-1]$ and $\rho_{\rm cs}$ being the initial density matrix of the central system, and $\mathcal{W}\left( {{t_2},{t_1}} \right)\rho = W(t_2,t_1)\rho W^{\dagger}(t_2,t_1)$ with $W(t_2,t_1)=\mathcal{T}\mathrm{exp}
(-i\int_{t_1}^{t_2} H_{\mathrm{eff}}(t')dt')$ being the no-jump propagator. $\mathcal{G}_p(t,0)$ contains all the paths with any sequence of $p$ ancilla jump events, therefore describing the $p$th-order ancilla errors. When $\kappa_lt,\gamma_jt\ll1$, the Liouville superoperator is well approximated by a finite-order Dyson expansion. 


\textit{Definition of path independence}.---The PI gates in this letter can be understood as follows. With an initial ancilla eigenstate $|i\rangle$ of $H_{\rm as}$, some control Hamiltonian acting during $[0,t]$ and a final projective measurement on the ancilla with result $|r\rangle$, the central system undergoes a deterministic unitary evolution up to finite-order or infinite-order Dyson expansion in Eq. (\ref{Dyson}). Now we provide a formal definition of path independence.

\textbf{Definition 1} (Path independence).---Let the ancilla start from $|i\rangle$ and end in $|r\rangle$, with $|i\rangle, |r\rangle\in\{|m\rangle\}_{m=0}^{d-1}$. Suppose
\begin{eqnarray}\label{df-pi}
  \langle r|\left[\sum_{p=0}^{k}\mathcal{G}_p(t,0)\left(|i\rangle\langle i|\otimes\rho_{\rm cs}\right)\right]|r\rangle\propto \mathcal{U}_{ri}(t,0)\rho_{\rm cs},
\end{eqnarray}
applies for $k\leq n$ but does not hold for $k>n$, where $\mathcal{U}_{ri}(t,0)\rho_{\rm cs}=U_{ri}(t,0)\rho_{\rm cs}U^{\dagger}_{ri}(t,0)$ is a unitary channel on the central system. Then we say the central system gate is PI of the ancilla errors up to the $n$th-order from $|i\rangle$ to $|r\rangle$.

\textit{Path independence for ancilla dephasing errors}.---The path independence for ancilla dephasing errors is guaranteed if the no-jump propagator is in a PI form below.

\textbf{Lemma 1}.---Let $\{U_{mn}(t_2,t_1)\}_{m,n=0}^{d-1}$ be a set of unitaries on the central system that are differentiable with respect to $t_2$ and $t_1$ and also satisfy the PI condition
\begin{eqnarray}\label{Upi}
 U_{me}(t_3,t_2)U_{en}(t_2,t_1)=U_{mn}(t_3,t_1),
\end{eqnarray}
with $m,e,n\in[0,d-1]$, there exist a class of PI no-jump propagators
\begin{eqnarray}\label{W}
 W(t_2,t_1)=\sum_{m,n}\xi_{mn}(t_2,t_1)|m\rangle\langle n|\otimes U_{mn}(t_2,t_1),
\end{eqnarray}
where $\{\xi_{mn}(t_2,t_1)\}_{m,n=0}^{d-1}$ are a set of complex functions of $t_2$ and $t_1$ satisfying $\xi_{mn}(t_3,t_1)=\sum_{e=0}^{d-1}\xi_{me}(t_3,t_2)\xi_{en}(t_2,t_1)$ and $\xi_{mn}(t,t)=\delta_{mn}$.

Note that here we define all the unitaries in the set $\{U_{mn}(t_2,t_1)\}_{m,n=0}^{d-1}$, but typically only a subset of $\{U_{mn}(t_2,t_1)\}_{m,n=0}^{d-1}$ with $\xi_{mn}(t_2,t_1)\neq0$ contribute to the no-jump dynamics and the other unitaries in the set with $\xi_{mn}(t_2,t_1)=0$ can be left undefined.

\textbf{Lemma 2}.---The PI condition for $\{U_{mn}(t_2,t_1)\}_{m,n=0}^{d-1}$ in Eq. (\ref{Upi}) is satisfied if and only if
\begin{eqnarray}\label{UmnR}
 U_{mn}(t_2,t_1)=R_m(t_2)U_{mn}R_n^{\dagger}(t_1),
\end{eqnarray}
where $R_m(t)=\mathcal{T}\{e^{-i\int_{0}^{t}H_m(t')dt'}\}$ with $H_m(t)$ being an arbitrary time-dependent Hamiltonian on the central system and $\{U_{mn}\}_{m,n=0}^{d-1}=\{U_{mn}(0,0)\}_{m,n=0}^{d-1}$ satisfy \cite{footnote2}
\begin{eqnarray}\label{loop}
 U_{me}U_{en}=U_{mn}.
\end{eqnarray}

\textbf{Theorem 1} (Dephasing errors).---With the PI no-jump propagator in Eq. (\ref{W}) and only ancilla dephasing errors, the central system gate is PI of all ancilla dephasing errors up to infinite order from $|i\rangle$ to $|r\rangle$ for all $|i\rangle,|r\rangle\in\{|m\rangle\}_{m=0}^{d-1}$.

\begin{figure}
\includegraphics[width=3.5in]{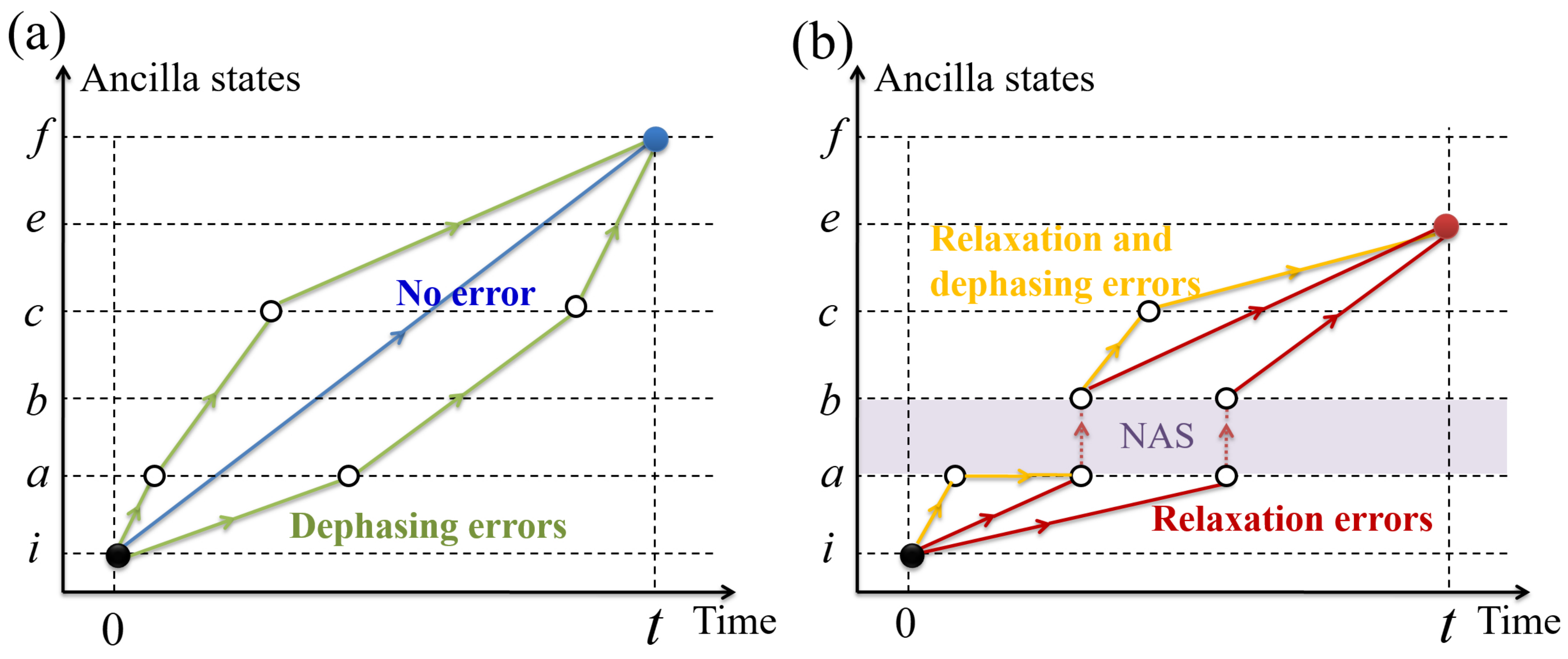}
\caption{
Schematic of ancilla evolution paths with different kinds of ancilla errors. (a) In the paths with blue line, the ancilla goes from $|i\rangle$ to $|f\rangle$ without any ancilla error and the central system gate is $U_{fi}$; In the paths with green lines, the ancilla suffers two dephasing errors $|a\rangle\langle a|$ and at $|c\rangle\langle c|$, while the central system gate is still $U_{fc}U_{ca}U_{ai}=U_{fi}$, independent of the dephasing error times. (b) In the paths with red lines, the ancilla sufferes a relaxation error $|b\rangle\langle a|$ (dashed red arrow lines) in the NAS spanned by $\{|a\rangle, |b\rangle\}$ (purple-shaded region) with the unitary gate as $U_{eb}U_{ai}$; In the paths with yellow line, the ancilla suffers two additional dephasing errors $|a\rangle\langle a|$ and $|c\rangle\langle c|$ but with the same unitary gate as that for a single relaxation error. The solid (hollow) circles represent the initial and final (intermediate) ancilla states, and the red dashed arrows represent the ancilla relaxation errors. Here we adopt the interaction picture associated with $H'_0(t)$.}
\label{path}
\end{figure}

To understand Theorem 1, we move to the the interaction picture associated with $H'_0(t)=\sum_{m=0}^{d-1}|m\rangle\langle m|\otimes H_m(t)$ [note that $H_0$ in Eq. (\ref{H0}) and $H'_0(t)$ are similar but can be different]. The no-jump propagator becomes
\begin{eqnarray}\label{WI}
    W^{(I)}(t_2,t_1)=\sum_{m,n}\xi_{mn}(t_2,t_1)|m\rangle\langle n|\otimes U_{mn},
\end{eqnarray}
and the ancilla dephasing operator $L_l^{(I)}(t_1)=L_l$ acts trivially on the central system regardless of the jump time $t_1$. Suppose the ancilla suffers a dephasing error $L^{(I)}(t_1)$ at time $t_1\in[0,t]$, the quantum operation on the central system is
\begin{equation}
\begin{split}\label{}
 \langle r|W^{(I)}(t,t_1)L^{(I)}(t_1)W^{(I)}(t_1,0)|i\rangle
 \propto U_{ri},
\end{split}
\end{equation}
where we have used Eq. (\ref{loop}). So independent of the error time $t_1$, the central system undergoes the same unitary gate as that without any ancilla error [$\langle r|W^{(I)}(t,0)|i\rangle\propto U_{ri}$]. The conclusion holds for arbitrary number of dephasing jumps during the gate, since $U_{ri}=U_{re}\cdots U_{ba}U_{ai}$ with $a,b,\cdots,e\in[0,d-1]$ from Eq. (\ref{loop}). An intuitive picture is provided in Fig. \ref{path}{\color{blue}(a)}. Without ancilla errors [blue path in Fig. \ref{path}{\color{blue}(a)}], the ancilla goes directly from the initial state to the final state with the target central system gate. With ancilla dephasing errors [green paths in Fig. \ref{path}{\color{blue}(a)}], the ancilla takes different continuous paths between the same initial and final states, but the central system gate remains unchanged since it depends only on the initial and final ancilla states.

\textit{Path independence for ancilla relaxation errors}.---The path independence for ancilla relaxation errors is slightly more demanding than that for ancilla dephasing errors. To see this, suppose the ancilla suffers a relaxation error $J^{(I)}(t_1)=|m\rangle\langle n|\otimes R^{\dagger}_{m}(t_1)R_{n}(t_1)$ at time $t_1$, the final quantum operation on the central system is
\begin{equation}\label{}
\begin{split}\label{Re}
 &\langle r|W^{(I)}(t,t_1)J^{(I)}(t_1)W^{(I)}(t_1,0)|i\rangle \\
 &\propto U_{rm}R^{\dagger}_{m}(t_1)R_{n}(t_1)U_{ni},
\end{split}
\end{equation}
which is typically a unitary gate depending on $t_1$, causing decoherence of the central system when averaged over $t_1$. This can be avoided if $R^{\dagger}_{m}(t_1)R_{n}(t_1)\propto e^{i\Delta\lambda t_1}$ or $H_{m}(t)-H_{n}(t)=\Delta\lambda$ with $\Delta\lambda\in\mathbb{R}$. So the following definition is motivated.

\textbf{Definition 2} (Noiseless ancilla subspace).---Denote the ancilla subspace spanned by $\{|k\rangle\}\subset\{|m\rangle\}_{m=0}^{d-1}$ satisfying $H_{m}(t)+\lambda_m(t)=H_{n}(t)+\lambda_n(t)$ with $\lambda_m(t),\lambda_n(t)\in \mathbb{R}$ for all $|m\rangle,|n\rangle\in\{|k\rangle\}$ as the \textit{noiseless ancilla subspace} (NAS).

Path independence of first-order ancilla relaxation errors is guaranteed if the same unitary operation is applied to the central system for all possible paths from $|i\rangle$ to $|r\rangle$ with at most one ancilla relaxation jump. For example, if $\xi_{ri}\neq0$ and there are no paths with first-order relaxation errors from $|i\rangle$ to $|r\rangle$, the the central system gate is still $U_{ri}$, or if $\xi_{ri}=0$ and the only path from $|i\rangle$ to $|r\rangle$ is through a relaxation operator $J^{(I)}$ in the NAS, then the central system gate is $U_{rm}U_{ni}$ (This is equivalent to redefining $U_{ri}=U_{rm}U_{mn}U_{ni}$ by setting $U_{mn}=\mathbb{I}$, since $U_{ri}$ in Eq. (\ref{WI}) is not well defined if $\xi_{ri}=0$). The conclusion can be extended to the cases for higher-order relaxation errors.

\textbf{Theorem 2} (Relaxation \& Dephasing errors).---With the PI no-jump propagator in Eq. (\ref{W}) and both ancilla relaxation and dephasing errors, if all the possible paths from $|i\rangle$ to $|r\rangle$ with at most $n$ sequential ancilla relaxation jumps, only include either the path without relaxation errors or the paths consisting of no more than $n$ sequential ancilla relaxation jumps in the NAS, and these paths produce the same unitary gate on the central system, which does not hold for all the paths from $|i\rangle$ to $|r\rangle$ with at most $(n+1)$ sequential ancilla relaxation jumps, then the central system gate is PI of the combination of up to the $n$th-order ancilla relaxation errors and up to infinite-order ancilla dephasing errors from $|i\rangle$ to $|r\rangle$.

Theorem 2 can be intuitively understood by the diagrams in Fig. \ref{path}{\color{blue}(b)}. With only ancilla relaxation errors in the NAS [red paths in Fig. \ref{path}{\color{blue}(b)}], the ancilla path is composed of discontinuous segments connected by the relaxation error operators, and the final unitary gate on the central system is often different from that without ancilla errors, but if the ancilla ends in other states, the central system still undergoes a deterministic unitary evolution. With both ancilla relaxation errors in the NAS and dephasing errors [orange paths in Fig. \ref{path}{\color{blue}(b)}], for each path segment connected by the relaxation errors, the ancilla goes another continuous way with the same initial and final states, so the final unitary gate on the central system is the same as that with only relaxation errors.

A special case of PI gates is the error-transparent gates, theoretically proposed \cite{Vy2013,Kapit2018} and recently experimentally demonstrated \cite{Ma2019} against a specific system error, with the error syndromes corresponding to the ancilla states here. The error transparency requires the physical Hamiltonian to commute with the errors when acting on the QEC code subspace \cite{footnote5}, corresponding to a PI no-jump propagator [Eq. (\ref{W})] with $\xi_{mn}=0$ for $m\neq n$ and all the ancilla errors are relaxation errors $|m\rangle\langle n|$ in the NAS, and thus fulfill the PI criterion. However, the PI gates contain a larger set of operations, because the PI criterion can be fulfilled with non-error-transparent Hamiltonians (see Supplementary Information \cite{SI} for general construction of the PI control Hamiltonian and jump operators).


\textit{Example: PI gates in superconducting circuits}.---We consider the implementation of the photon-number selective arbitrary phase (SNAP) gates in superconducting circuits \cite{Krastanov2015,Heeres2015}.
The superconducting cavity (central system) dispersively couples to a nonlinear transmon device (ancilla system) with Hamiltonian
$H_{0}=\omega_{ge}|e\rangle\langle e|+\omega_{\rm c}a^{\dagger}a -\chi a^{\dagger}a|e\rangle\langle e|$ \cite{Schuster2007}, where $\omega_{ge}$ ($\omega_{\rm c}$) are the transmon (cavity) frequency, $a$ ($a^{\dagger}$) is the annihilation (creation) operator of the cavity mode, $\chi$ is the dispersive coupling strength, and $|e\rangle$ ($|g\rangle$) denotes the excited (ground) state of the ancilla transmon. The SNAP gate on the cavity, $S(\vec \varphi)=\sum_{n =0}^{\infty} e^{i\varphi_n}|n\rangle\langle n|$, imparts arbitrary phases $\vec \varphi=\{\varphi_n\}_{n=0}^{\infty}$ to the different Fock states of the cavity.

In the interaction picture associated with $H_0$, the PI control Hamiltonian is $H_{\rm c}^{(I)}=\Omega[|g\rangle\langle e|\otimes S(\vec \varphi)+|e\rangle\langle g|\otimes S(-\vec \varphi)]+\delta|e\rangle\langle e|=\Omega\sum_{n =0}^{\infty}( e^{i\varphi_n}|g,n\rangle\langle e,n|+e^{-i\varphi_n}|e,n\rangle\langle g,n|)+\delta|e\rangle\langle e|$ \cite{footnote4}, inducing a PI propagator [Eq. (\ref{WI})] \cite{SI}. Returning to the Schr\"{o}dinger's picture, $H_{\rm c}(t)=\Omega\sum_{n =0}^{\infty}( e^{i[(\omega_{ge}-n\chi)t+\varphi_n-\delta]}|g,n\rangle\langle e,n|+\rm H.c.)$. When $\Omega\ll\chi$, the control Hamiltonian can be simplified as the driving acting on the transmon alone but with multiple frequency components, $H_{\rm c}(t)\approx \epsilon_{ge}(t)e^{i(\omega_{ge}-\delta)t}|g\rangle\langle e|+\text{H.c.}$ with $\epsilon_{ge}(t)=\sum_n \Omega e^{i(\varphi_n-n\chi t)}$. Then the SNAP gates are PI of any transmon dephasing error with $\mathcal{D}[\sqrt{\gamma_1}(c_e|e\rangle\langle e|+c_g|g\rangle\langle g|)]$ with $c_{g/e}\in\mathbb{C}$ (see Theorem 1), but not PI of the transmon relaxation error with $\mathcal{D}[\sqrt{\kappa_1}(|g\rangle\langle e|)]$ since $\{|g\rangle, |e\rangle\}$ do not span a NAS (Theorem 2) \cite{SI}.

To make the SNAP gates PI of the dominant transmon relaxation error, we use a 3-level transmon with $H_{0}=\omega_{ge}|e\rangle\langle e|+\omega_{gf}|f\rangle\langle f|+\omega_{\rm c}a^{\dagger}a -\chi (|e\rangle\langle e|+|f\rangle\langle f|)a^{\dagger}a$, where the dispersive coupling strength is engineered to the same for the first-excited transmon state $|e\rangle$ and the second-excited state $|f\rangle$ \cite{Rosenblum2018}. The SNAP gate is implemented by applying the Hamiltonian that drives the $|g\rangle\leftrightarrow|f\rangle$ transition instead of the $|g\rangle\leftrightarrow|e\rangle$ transition \cite{footnote1}.
Since $\{|e\rangle, |f\rangle\}$ span a NAS, the SNAP gate is PI of the dominant transmon relaxation error with $\mathcal{D}[\sqrt{\kappa_2}(|e\rangle\langle f|)]$ and also of any transmon dephasing error with $\mathcal{D}[\sqrt{\gamma_2}(c_{g}|g\rangle\langle g|+c_e|e\rangle\langle e|+c_f|f\rangle\langle f|)]$. Note that the PI SNAP gates are not error-transparent, but still enable robustness against transmon errors.


\textit{PI gates for both ancilla errors and central system errors}.---The PI gates for ancilla errors can also be made PI of the central system errors. We assume that the central system also suffers Markovian noise with the Lindbladian dissipators $\sum_{i=0}^{q-1}\mathcal{D}[\sqrt{\zeta}_i E_i]$. Suppose a quantum error correction (QEC) code exists for the central system \cite{Nielsen2010,Shor1995,Knill1997}, which means that the error set $\{E_i\}$ satisfy the Knill-Laflamme condition $P_0E_i^{\dagger}E_jP_0=A_{ij}P_0$ with $E_0=P_0$ being the projection to the code subspace and $A$ a Hermitian matrix. We may diagonalize $A$ as $B=u^{\dagger}Au$ to obtain another set of correctable errors $\{F_k\}$ with $F_k=\sum_{ik}u_{ik}E_i$, satisfying $P_0F_k^{\dagger}F_lP_0=r_{k}\delta_{kl}P_0$ with $F_0=P_0$, $B_{00}=1$ and $r_k=B_{kk}$. Then the condition for path independence against the central system errors is
\begin{eqnarray}\label{}
  [H'_{0}(t),F_k]=\sum_{m=0}^{d-1}c_{m,k}(t)|m\rangle\langle m|\otimes F_k,
\end{eqnarray}
where $m\in[0,d-1],~k\in[0,q-1]$ and $c_{m,k}(t)\in\mathbb{R}$. In the interaction picture associated with $H'_0(t)$, this condition ensures that $R^{\dagger}(t)F_kR(t) =\sum_m e^{i\int_{t_1}^{t_2} c_{m,k}(t')dt'}|m\rangle\langle m|\otimes F_k$ is a tensor product of the ancilla dephasing operator and the same error operator $F_k$ with $R(t)=\mathcal{T}e^{-i\int_{0}^{t} H'_0(t')dt'}$. Then the PI no-jump propagator for both ancilla and central system errors \cite{Vy2013,Kapit2018,SI} can be constructed as in Eq. (\ref{WI}) with
\begin{eqnarray}\label{FTc}
  U_{mn}=\sum_{k}e^{i\phi_{mn,k}}F_kU_{mn,0} F_k^{\dagger}/r_{k},
\end{eqnarray}
where $U_{mn,0}$ is the target unitary in the code subspace satisfying $U_{mn,0}U_{mn,0}^{\dagger}=P_0$ and $\phi_{mn,k}\in\mathbb{R}$. After such a PI gate, we can make a joint measurement on both the ancilla state and the error syndromes of the central system, then the path independence of ancilla errors is ensured and the first-order central system errors during the gate can also be corrected \cite{SI}.

\textit{Summary}.---To address the challenge of ancilla-induced decoherence, we provide a general criterion of path independence. For quantum information processing with bosonic encoding, such a PI design will be crucial in protecting the encoded information from ancilla errors, while the previous transversal approach does not apply. Moreover, different from the traditional approaches with separated quantum control and error correction tasks, our approach integrates quantum control and error correction. Using the general PI design, we can further explore PI gates using various kinds of ancilla systems to achieve higher-order suppression of ancilla errors, design PI operations robust against both ancilla errors and central system errors, and extend the PI technique to quantum sensing and other quantum information processing tasks.


We thank Chang-Ling Zou, Stefan Krastanov, Changchun Zhong, Sisi Zhou, Connor Hann, Linshu Li and Chiao-Hsuan Wang for helpful discussions.
We acknowledge support from the ARO (W911NF-18-1-0020, W911NF-18-1-0212), ARL-CDQI (W911NF-15-2-0067), ARO MURI (W911NF-16-1-0349 ), AFOSR MURI (FA9550-15-1-0015, FA9550-19-1-0399), DOE (DE-SC0019406), NSF (EFMA-1640959, OMA-1936118), and the Packard Foundation (2013-39273).

\textit{Note added}.---Recently the PI SNAP gates have been experimentally implemented in a superconducting circuits \cite{Rosenblum2019}, with the SNAP gate fidelity significantly improved by the PI design.



\begin{thebibliography}{}
\bibitem{footnote3} There is no basic obstacle to extend the mathematical framework in this paper to the case with an
infinite-dimensional ancilla or a continuous variable ancilla system, as long as a countable projection eigenbasis can be identified for the ancilla system. But for ancilla-assisted control, it is often sufficient to consider a finite-dimensional subspace of the infinite-dimensional ancilla.
\bibitem{footnote2} This is analogous to the cocycle condition that the transition functions of the coordinates of any two overlapping patches in a manifold should satisfy.
\bibitem{footnote5} A more general condition for error transparency is that the commutator of the physical Hamiltonian and the error operator when acting on the QEC code subspace is proportional to the error operator.
\bibitem{footnote4} The original SNAP gate consists of two consequtive $\pi$-pulses (between $|g,n\rangle$ and $|e,n\rangle$) with a geometric phase depending on the phase difference between the two $\pi$ pulses. For simplicity, we may fix the phase of the second pulse to be 0, so that the geometric phase for the SNAP gate is determined by the phase $\varphi_n$ of the first $\pi$ pulse.
\bibitem{footnote1} In practice, $|g\rangle\leftrightarrow|f\rangle$ transition is not dipole-allowed but can be realized by a Raman drive with largely detuned $|g\rangle\leftrightarrow|e\rangle$ and $|e\rangle\leftrightarrow|f\rangle$ transitions.
\end{thebibliography}
\end{document}


\title{Supplementary Information for ``Path-Independent Quantum Gates with Noisy Ancilla''}
\author{Wen-Long Ma}
\affiliation{Pritzker School of Molecular Engineering, University of Chicago, Illinois 60637, USA}
\affiliation{Department of Applied Physics and Physics, Yale University, New Haven, Connecticut 06511, USA}
\affiliation{Yale Quantum Institute, Yale University, New Haven, Connecticut 06511, USA}
\author{Mengzhen Zhang}
\affiliation{Pritzker School of Molecular Engineering, University of Chicago, Illinois 60637, USA}
\affiliation{Department of Applied Physics and Physics, Yale University, New Haven, Connecticut 06511, USA}
\affiliation{Yale Quantum Institute, Yale University, New Haven, Connecticut 06511, USA}
\author{Yat Wong}
\affiliation{Pritzker School of Molecular Engineering, University of Chicago, Illinois 60637, USA}
\author{Kyungjoo Noh}
\affiliation{Pritzker School of Molecular Engineering, University of Chicago, Illinois 60637, USA}
\affiliation{Department of Applied Physics and Physics, Yale University, New Haven, Connecticut 06511, USA}
\affiliation{Yale Quantum Institute, Yale University, New Haven, Connecticut 06511, USA}
\author{Serge~Rosenblum}
\affiliation{Department of Applied Physics and Physics, Yale University, New Haven, Connecticut 06511, USA}
\affiliation{Yale Quantum Institute, Yale University, New Haven, Connecticut 06511, USA}
\affiliation{Department of Condensed Matter Physics, Weizmann Institute of Science, Rehovot, Israel}
\author{Philip Reinhold}
\affiliation{Department of Applied Physics and Physics, Yale University, New Haven, Connecticut 06511, USA}
\affiliation{Yale Quantum Institute, Yale University, New Haven, Connecticut 06511, USA}
\author{Robert J. Schoelkopf}
\affiliation{Department of Applied Physics and Physics, Yale University, New Haven, Connecticut 06511, USA}
\affiliation{Yale Quantum Institute, Yale University, New Haven, Connecticut 06511, USA}
\author{Liang Jiang}
\affiliation{Pritzker School of Molecular Engineering, University of Chicago, Illinois 60637, USA}
\affiliation{Department of Applied Physics and Physics, Yale University, New Haven, Connecticut 06511, USA}
\affiliation{Yale Quantum Institute, Yale University, New Haven, Connecticut 06511, USA}

\date{\today }

\begin{abstract}

\end{abstract}

\maketitle


\maketitle

\tableofcontents
\section{Proof of Lemma 1}
\textbf{Lemma 1}.---Let $\{U_{mn}(t_2,t_1)\}_{m,n=0}^{d-1}$ be a set of unitaries on the central system that are differentiable with respect to $t_2$ and $t_1$ and also satisfy the PI condition
\begin{align}\label{Upi}
 U_{me}(t_3,t_2)U_{en}(t_2,t_1)=U_{mn}(t_3,t_1),
\end{align}
with $m,e,n\in[0,d-1]$, there exist a class of path-independent (PI) no-jump propagators
\begin{align}\label{W}
 W(t_2,t_1)=\sum_{m,n}\xi_{mn}(t_2,t_1)|m\rangle\langle n|\otimes U_{mn}(t_2,t_1),
\end{align}
where $\{\xi_{mn}(t_2,t_1)\}_{m,n=0}^{d-1}$ are a set of complex functions of $t_2$ and $t_1$ satisfying $\xi_{mn}(t_3,t_1)=\sum_{e=0}^{d-1}\xi_{me}(t_3,t_2)\xi_{en}(t_2,t_1)$ and $\xi_{mn}(t,t)=\delta_{mn}$.

\begin{proof}
---The no-jump propagator is defined in the main text as
\begin{align}\label{}
 W(t_2,t_1)=\mathcal{T}\mathrm{exp}\left(-i\int_{t_1}^{t_2} H_{\mathrm{eff}}(t')dt'\right),
\end{align}
so it should satisfy
\begin{align}\label{decom}
    W(t_3,t_1)=W(t_3,t_2)W(t_2,t_1),
\end{align}
implying that any matrix element between the ancilla states for both sides of the equation should be the same,
\begin{align}\label{}
 \langle m|W(t_3,t_1)|n\rangle=&\langle m|W(t_3,t_2)W(t_2,t_1)|n\rangle, \nonumber \\
 =&\sum_{e=0}^{d-1}\xi_{me}(t_{3},t_{2})\xi_{en}(t_2,t_1)U_{me}(t_3,t_2)U_{en}(t_2,t_1), \nonumber \\
 =&\sum_{e=0}^{d-1}\xi_{me}(t_{3},t_{2})\xi_{en}(t_2,t_1)U_{mn}(t_3,t_1),
\end{align}
where we have used $U_{me}(t_3,t_2)U_{en}(t_2,t_1)=U_{mn}(t_3,t_1)$ from the second line to the third line. So $W(t_3,t_1)$ is still in the same form as Eq. (\ref{W}) with $\{U_{mn}(t_2,t_1)\}$ satisfying the PI condition. If we require $\xi_{mn}(t_2,t_1)$ to be a well-defined function of $t_2$ and $t_1$, then $\xi_{mn}(t_2,t_1)$ should satisfy
\begin{align}
 \xi_{mn}(t_3,t_1)=\sum_{e=0}^{d-1}\xi_{me}(t_3,t_2)\xi_{en}(t_2,t_1).
\end{align}

The no-jump propagator should also satisfy that $W(t,t)=I$ with $I$ being the identity operator for the whole system, which requires that $\xi_{mn}(t,t)=\delta_{mn}$ with $\delta_{mn}$ being the Kronecker delta function. The existence of $\{\xi_{mn}(t_2,t_1)\}$ will be verified by constructing an explicit example in Sec. \ref{exact}. Moreover, we have $W(t_2,t_2)=W(t_2,t_1)W(t_1,t_2)=I$, so $W(t_2,t_1)^{-1}=W(t_1,t_2)$.
\end{proof}

Note that here we define all the unitaries in the set $\{U_{mn}(t_2,t_1)\}_{m,n=0}^{d-1}$, which satisfy the PI condition in Eq. (\ref{Upi}), but in practice only a subset of $\{U_{mn}(t_2,t_1)\}_{m,n=0}^{d-1}$ with $\xi_{mn}(t_2,t_1)\neq0$ contribute to the no-jump dynamics. In this case, to have a PI no-jump propagator as that in Eq. (\ref{W}), we only need to define the unitaries in such a subset to satisfy the PI condition, and the other unitaries in the set $\{U_{mn}(t_2,t_1)\}_{m,n=0}^{d-1}$ with $\xi_{mn}(t_2,t_1)=0$ can be left undefined.



\section{Proof of Lemma 2}
\textbf{Lemma 2}.---The PI condition for $\{U_{mn}(t_2,t_1)\}_{m,n=0}^{d-1}$ in Eq. (\ref{Upi}) is satisfied if and only if
\begin{align}\label{UmnR}
 U_{mn}(t_2,t_1)=R_m(t_2)U_{mn}R_n^{\dagger}(t_1),
\end{align}
where $R_m(t)=\mathcal{T}\{e^{-i\int_{0}^{t}H_m(t')dt'}\}$ with $H_m(t)$ being an arbitrary time-dependent Hamiltonian on the central system and $\{U_{mn}\}_{m,n=0}^{d-1}=\{U_{mn}(0,0)\}_{m,n=0}^{d-1}$ satisfy
\begin{align}\label{loop}
 U_{me}U_{en}=U_{mn}.
\end{align}

\begin{proof}
--- The `if' part is easy to prove. Suppose $U_{mn}(t_2,t_1)=R_m(t_2)U_{mn}R_n^{\dagger}(t_1)$ with $\{U_{mn}\}_{m,n=0}^{d-1}$ satisfying the condition in Eq. (\ref{loop}), then
\begin{align}\label{}
 &U_{me}(t_3,t_2)U_{en}(t_2,t_1)=R_m(t_3)U_{me}R_e^{\dagger}(t_2)R_e(t_2)U_{en}R_n(t_1) \nonumber \\
 =&R_m(t_3)U_{me}U_{en}R_n(t_1)=R_m(t_3)U_{mn}R_n(t_1)=U_{mn}(t_3,t_1).
\end{align}

Conversely, suppose $U_{me}(t_3,t_2)U_{en}(t_2,t_1)=U_{mn}(t_3,t_1)$, then
\begin{align}\label{Udt}
   U_{me}(t_3,t_2+\Delta t)U_{en}(t_2+\Delta t,t_1)=U_{mn}(t_3,t_1).
\end{align}
Since we assume that $\{U_{mn}(t_2,t_1)\}$ are differentiable with respect to both $t_1$ and $t_2$, so we can define
\begin{align}\label{}
 &\frac{\partial U_{mn}(t_2,t_1)}{\partial t_2}=-iH^{(l)}_{mn}(t_2,t_1)U_{mn}(t_2,t_1), \\
 &\frac{\partial U_{mn}(t_2,t_1)}{\partial t_1}=iU_{mn}(t_2,t_1)H^{(r)}_{mn}(t_2,t_1),
\end{align}
so we have
\begin{align}\label{}
  &U_{me}(t_3,t_2+\Delta t)U_{en}(t_2+\Delta t,t_1) \nonumber \\
  =&\left[U_{me}(t_3,t_2)-i\Delta t U_{me}(t_3,t_2)H^{(r)}_{me}(t_3,t_2)\right]\left[U_{en}(t_2,t_1)+i\Delta t H^{(l)}_{en}(t_2,t_1) U_{en}(t_2,t_1)\right] \nonumber \\
  =&U_{me}(t_3,t_2)U_{en}(t_2,t_1)-i\Delta tU_{me}(t_3,t_2)[H^{(r)}_{me}(t_3,t_2)-H^{(l)}_{en}(t_2,t_1)]U_{en}(t_2,t_1)+\mathcal{O}({\Delta t}^2).
\end{align}
Eq. (\ref{Udt}) requires that $H^{(r)}_{me}(t_3,t_2)=H^{(l)}_{en}(t_2,t_1)$ for any $t_1$, $t_2$, $t_3$, and $m,e,n\in[0,d-1]$, implying that $H^{(r)}_{me}(t_3,t_2)$ is independent of $t_3$ and $m$ and $H^{(l)}_{en}(t_2,t_1)$ is independent of $t_1$ and $n$, so $H^{(r)}_{me}(t_3,t_2)=H^{(l)}_{en}(t_2,t_1)=H_e(t_2)$. Then $U_{mn}(t_2,t_1)$ can be obtained from $U_{mn}=U_{mn}(0,0)$ by first integrating the first variable from $0$ to $t_2$ and then the second variable from $0$ to $t_1$, so $U_{mn}(t_2,t_1)=R_m(t_2)U_{mn}R_n^{\dagger}(t_1)$ with $R_m(t)=\mathcal{T}\{e^{-i\int_{0}^{t}H_m(t')dt'}\}$. Inserting this expression of $U_{mn}(t_2,t_1)$ into $U_{me}(t_3,t_2)U_{en}(t_2,t_1)=U_{mn}(t_3,t_1)$, we have $U_{me}U_{en}=U_{mn}$. 
\end{proof}

From Eq. (\ref{loop}), we can derive that (i) $U_{mm}=\mathbb{I}$, (ii) $U_{mn}=U^{\dagger}_{nm}$; (iii) $U_{ma}=U_{me}\cdots U_{cb}U_{ba}$. So Eq. (\ref{loop}) is equivalent to
\begin{align}\label{loop2}
    U_{me}\cdots U_{cb}U_{ba}U_{am}=\mathbb{I},
\end{align}
with $m,a,b,c,\cdots,e\in[0,d-1]$. From the viewpoint of non-Abelian path integration (Table \ref{table}) \cite{Broda2001}, the set of discrete ancilla states $\{|m\rangle\}_{m=0}^{d-1}$ defines a manifold, $U_{mn}$ is the parallel-transport operator from $|n\rangle$ to $|m\rangle$, and then Eq. (\ref{loop2}) means that the holonomy for any loop path $|m\rangle\rightarrow|a\rangle\rightarrow|b\rangle\rightarrow|c\rangle\cdots\rightarrow|e\rangle\rightarrow|m\rangle$ is always the identity.

Now we can write the explicit form of PI no-jump propagator in Eq. (\ref{W}) as
\begin{align}\label{We}
  W(t_2,t_1)=\sum_{m=0}^{d-1}\xi_{mm}(t_2,t_1)|m\rangle\langle m|\otimes \mathbb{I}+\sum_{m\neq n}\xi_{mn}(t_2,t_1)|m\rangle\langle n|\otimes R_m(t_2)U_{mn}R_n^{\dagger}(t_1).
\end{align}
Note that if $\xi_{mn}(t_2,t_1)=0$ with $m,n\in[0,d-1]$ and $m\neq n$, then $U_{mn}$ is not well defined, but it is still possible to redefine $U_{mn}$ after considering the ancilla relaxation errors (see Proof of Theorem 2 below).

\newcommand{\tabincell}[2]{\begin{tabular}{@{}#1@{}}#2\end{tabular}}
\begin{table}[t]
\begin{spacing}{1.2}
\begin{tabular*}{12.5cm}{ccc}
\hline
\hline
\multirow{2}{*}
{} & \tabincell{c}{Non-Abelian path integration}  & PI propagator  \\
\hline
\tabincell{c}{Parrallel-transport operator} & $\exp(i\int_{L} \mathbf{A})$ & $U_{mn}$  \\
\hline
Holonomy  & $\exp(i\oint \mathbf{A})$ &  $U_{me}\cdots U_{cb}U_{ba}U_{am}$ \\
\hline
\hline
\end{tabular*}
\end{spacing}
\caption{Comparison between the non-Abelian path integration and the PI unitary propagator for central system. Here $\mathbf{A}$ is a connection one-form. }
\label{table}
\end{table}

\section{Proof of Theorem 1, Theorem 2 and examples}





\textbf{Theorem 1} (Dephasing errors).---With the PI no-jump propagator in Eq. (\ref{W}) and only ancilla dephasing errors, the central system gate is PI of all ancilla dephasing errors up to infinite order from $|i\rangle$ to $|r\rangle$ for all $|i\rangle,|r\rangle\in\{|m\rangle\}_{m=0}^{d-1}$.

\begin{proof}
---Consider a sequence of $n$ dephasing errors $L_l\rightarrow L_k\cdots\rightarrow L_q$ at times $0\leq t_1<t_2\cdots<t_n\leq t$, then the propagator for the central system can be expressed as the matrix element of the propagator for the whole system between the initial state $|i\rangle$ and final state $|l\rangle$ of the ancilla,
\begin{align}
 &\langle r|W(t,t_n)L_q\cdots L_k W(t_2,t_1)L_lW(t_1,0)|i\rangle \nonumber \\
 =&\sum_{a=0}^{d-1}\cdots\sum_{b=0}^{d-1}\sum_{c=0}^{d-1}\Delta_{q}^{(a)}\cdots \Delta_{k}^{(b)}\Delta_{l}^{(c)}\xi_{ra}(t,t_n)\cdots\xi_{bc}(t_2,t_1)
 \nonumber \\
 &\times \xi_{ci}(t_1,0)U_{ra}(t,t_n)\cdots U_{bc}(t_2,t_1)U_{ci}(t_1,0)  \nonumber \\
 \propto& U_{ri}(t,0),
\end{align}
where we have used Eq. (\ref{Upi}) in Lemma 1. Since the final unitary gate on the central system is independent of the dephasing error sequence $\{L_l,L_k,\cdots L_q\}$ and the error times $\{t_1,t_2,\cdots,t_n\}$, so the gate on the central system remains unchanged even after the path integral in Eq. {\color{blue}(8)} of the main text. Typically we should have $\xi_{ri}\neq0$, ensuring that the ancilla $|i\rangle\rightarrow|r\rangle$ transition is permitted by the PI no-jump propagator.
\end{proof}

\textbf{Theorem 2} (Relaxation \& Dephasing errors).---With the PI no-jump propagator in Eq. (\ref{W}) and both ancilla relaxation and dephasing errors, if all the possible paths from $|i\rangle$ to $|r\rangle$ with at most $n$ sequential ancilla relaxation jumps, only include either the path without relaxation errors or the paths consisting of no more than $n$ sequential ancilla relaxation jumps in the NAS, and these paths produce the same unitary gate on the central system, which does not hold for all the paths from $|i\rangle$ to $|r\rangle$ with at most $(n+1)$ sequential ancilla relaxation jumps, then the central system gate is PI of the combination of up to the $n$th-order ancilla relaxation errors and up to infinite-order ancilla dephasing errors from $|i\rangle$ to $|r\rangle$.

\begin{proof}
---First consider $\xi_{ri}\neq0$ and there are no other paths from $|i\rangle$ to $|r\rangle$ up to the $n$th-order ancilla relaxation errors, then the central system will undergoes a unitary gate $U_{ri}$ up to the $n$th-order ancilla relaxation errors, which is the same unitary gate as that without any ancilla errors.

If $\xi_{ri}=0$, and the only path from $|i\rangle$ to $|r\rangle$ up to the $n$th-order ancilla relaxation errors is through a single relaxation operator $J_j=|m_j\rangle\langle n_j|$, then the propagator for the central system is
\begin{align}
 &\langle r|W(t,t_1)J_jW(t_1,0)|i\rangle
 \propto U_{r,m_j}(t,t_1)U_{n_j,i}(t_1,0)\nonumber \\
 =&R_r(t)U_{r,m_l}R_{m_l}^{\dagger}(t_1)R_{n_l}(t_1)U_{n_l,i},
\end{align}
where we have used Eq. (\ref{UmnR}) in Lemma 2. If $|m_j\rangle$ and $|n_j\rangle$ are in the NAS with $H_{m_j}(t)$ and $H_{n_j}(t)$ differing only by some real constant, the central system undergoes the final unitary evolution $R_r(t)U_{r,m_j}U_{n_j,i}$ independent of the jump time $t_1$ (This is equivalent to redefining $U_{ri}=U_{r,m_j}U_{n_j,i}$ by setting $U_{m_j,n_j}=\mathbb{I}$, since $U_{ri}$ is not well defined if $\xi_{ri}=0$). Here $\xi_{ri}=0$ ensures that the ancilla relaxation path does not interfere with the path without errors.

If $\xi_{ri}=0$, and the only path from $|i\rangle$ to $|r\rangle$ up to the $n$th-order ancilla relaxation errors, the only path consists of $n$ sequential relaxation jumps $J_j\rightarrow J_k\cdots\rightarrow J_q$ in the NAS at times $0\leq t_1<t_2\cdots<t_n\leq t$, the final unitary operation on the central system becomes $R_r(t)U_{r,m_q}U_{n_q,.}\cdots U_{.,m_k}U_{n_k,m_j}U_{n_j,i}$ (redefining $U_{ri}=U_{r,m_q}U_{n_q,.}\cdots U_{.,m_k}U_{n_k,m_j}U_{n_j,i}$).

In practice, there may be more than one paths from $|i\rangle$ to $|r\rangle$ with at most $n$ sequential ancilla relaxation jumps, but if all these paths only include either the path without relaxation errors or the paths with no more that $n$ relaxation jumps in the NAS, and produce the same unitary operation on the central system, then the central system still undergoes a deterministic unitary gate. If the above conclusion holds for all the paths with at most $n$ sequential ancilla relaxation jumps, but not for all the paths with at most $(n+1)$ sequential ancilla relaxation jumps, the central system gate is PI of the relaxation errors up to the $n$th-order from $|i\rangle$ to $|r\rangle$ (according to Definition 1). Any infinite sequence of dephasing errors in any time interval (say $[t_1,t_2]$) leave the unitary operation for each time interval (e.g., $U_{n_k,m_j}$) unchanged, therefore the central system undergoes the same final unitary operation as that with only relaxation errors, so the central system gate is PI of the relaxation errors up to the combination of up to the $n$th-order ancilla relaxation errors and up to infinite-order ancilla dephasing errors from $|i\rangle$ to $|r\rangle$.
\end{proof}

In Theorem 1 and Theorem 2, we classify the ancilla errors into dephasing errors, relaxation errors, and the combination of both, and analyse the path independence for different types ancilla errors separately. With the ancilla starting from $|i\rangle$ and ending in $|r\rangle$, the central system gate is PI of up to infinite-order ancilla dephasing errors and up to the $n$th-order ancilla relaxation errors. In the presence of all three types of ancilla errors, the central system gate is PI of up to the $n$th-order ancilla errors, i.e. the gate fidelity from $|i\rangle$ to $|r\rangle$ is limited by the $(n+1)$th-order ancilla relaxation errors. Moreover, for a specific ancilla initial state $|i\rangle$, we need to analyse the path independence for all the possible ancilla final states $\{|r\rangle\}_{r=0}^{d-1}$, and the overall gate fidelity is limited by the worst case of all the possible final states.

\begin{figure}
\includegraphics[width=3.5in]{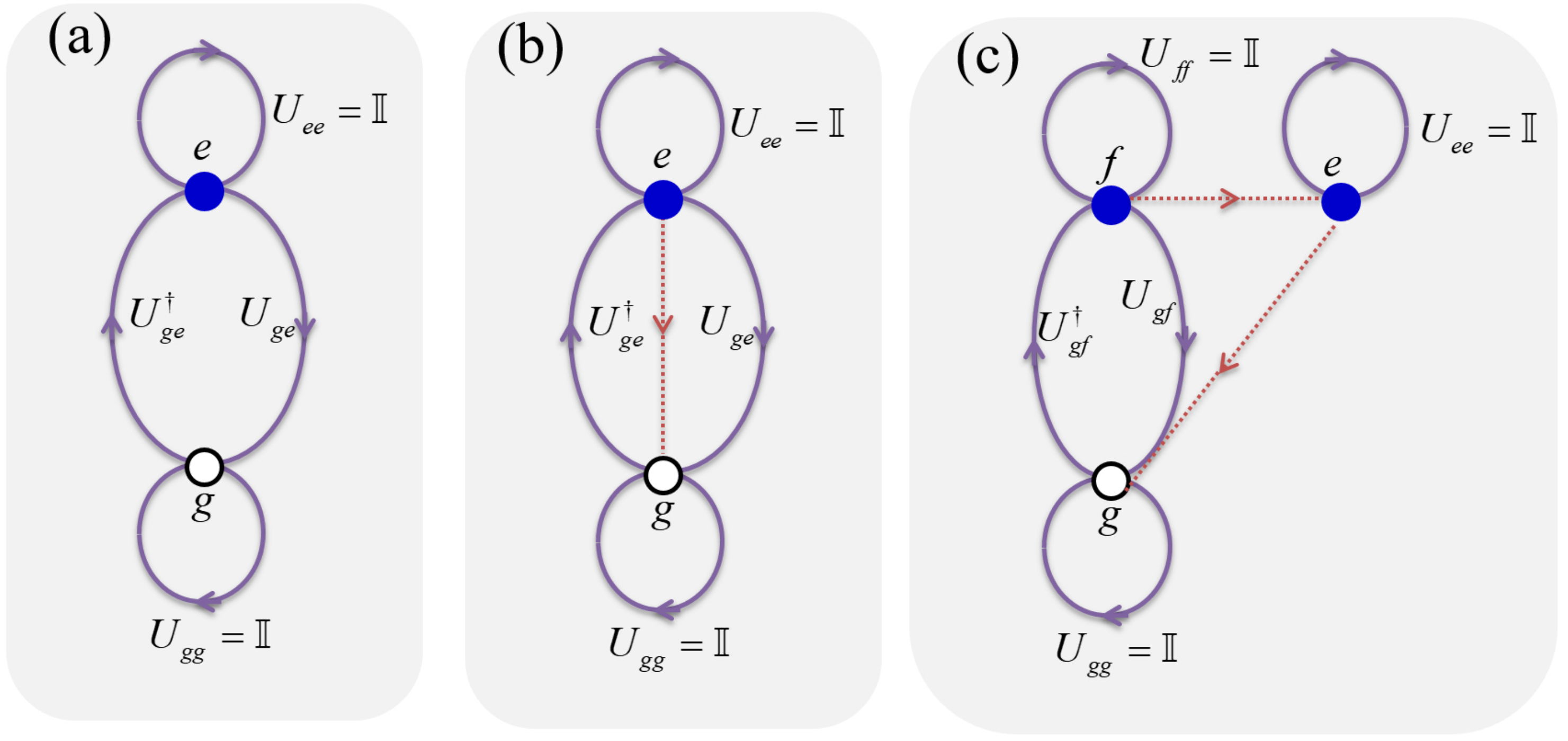}
\includegraphics[width=3.4in]{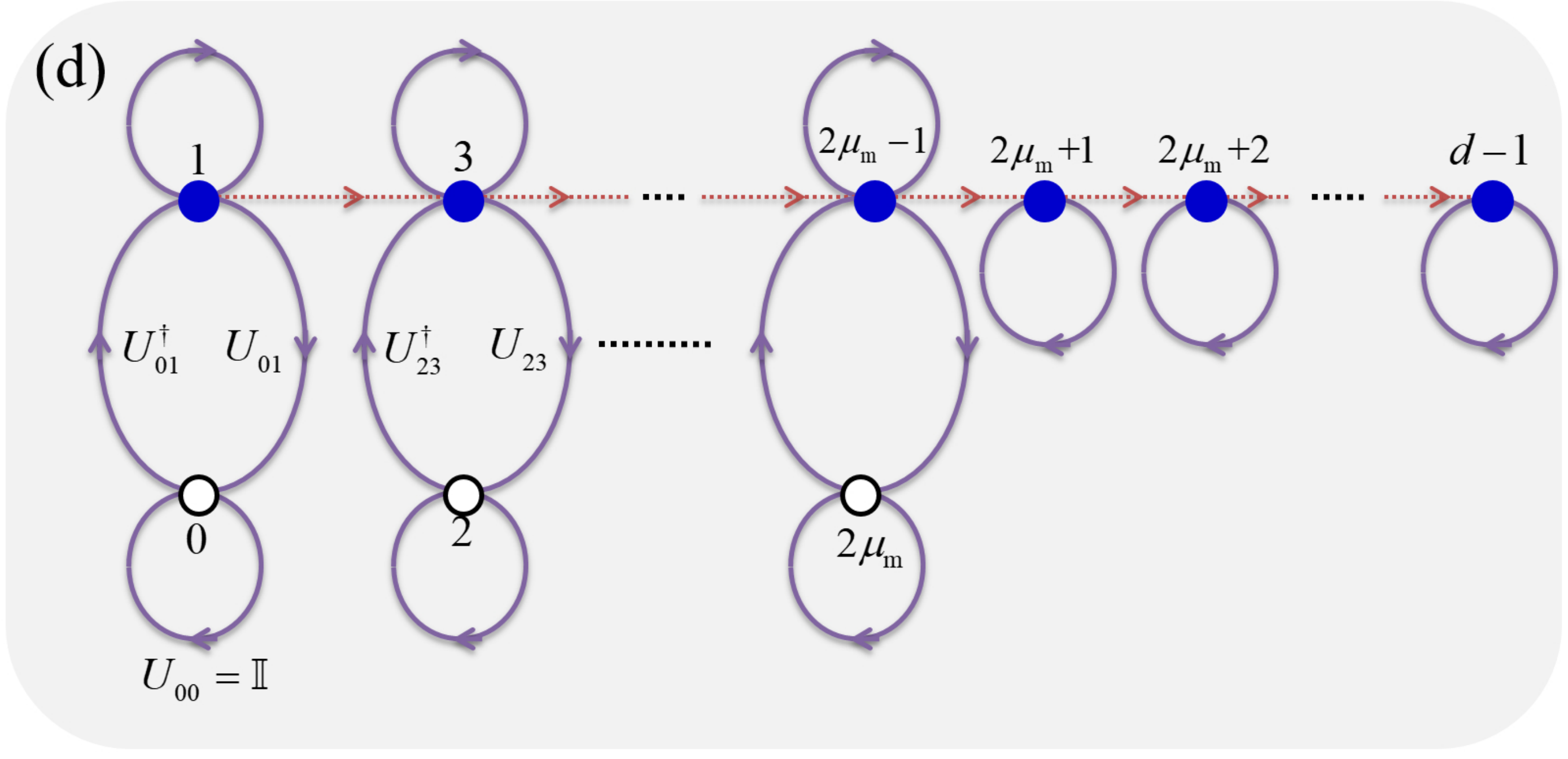}
\caption{Schematic of the PI control protocols for a 2-level, 3-level and $d$-level ancilla. (a) without ancilla relaxation errors, the central system gate assisted by a 2-level ancilla is PI of infinite-order dephasing errors. (b) with an ancilla relaxation error $|g\rangle\langle e|$ (not in the NAS), the PI condition is not fulfilled and the central system gate fidelity is limited by the first-order ancilla relaxation error. (c) For a 3-level ancilla with a NAS spanned by $\{|e\rangle, |f\rangle\}$, the central system gate is PI of the first-order ancilla relaxation error ($|e\rangle\langle f|$) and the gate fidelity is limited by the second-order ancilla relaxation error (a sequence of $|e\rangle\langle f|$ and $|g\rangle\langle e|$). (d) For a $d$-level ancilla with all the relaxation errors in the NAS, the central system gate is PI of any possible ancilla errors. In (a)-(d), the dephasing errors make the ancilla undergo additional 1-site or 2-site loops, while ancilla relaxation errors (represented by the red dashed arrows) connect the otherwise discontinous ancilla paths. The ancilla states in the NAS is represented by blue solid circles while the other ancilla states are represented by black hollow circles (Strictly speaking, the circles represent the central system conditioned on a specific ancilla state). Here we adopt the interaction picture associated with $H'_0(t)$.}
\label{level}
\end{figure}

\subsection{Examples}
To illustrate how to apply Theorem 1 and Theorem 2, we give several examples below, including 2-level, 3-level and $d$-level ancillas. The case of a 2-level ancilla corresponds to the SNAP gate on a cavity assisted by a 2-level transmon, while that of 3-level ancilla corresponds to the PI SNAP gate assisted by a 3-level transmon.

To simplify the notation, we move to the interaction picture associated with $H'_0(t)=\sum_{m=0}^{d-1}|m\rangle\langle m|\otimes H_m(t)$. Then the PI no-jump propagator in Eq. (\ref{We}) becomes
\begin{align}\label{WeI}
   W^{(I)}(t_2,t_1)=R^{\dagger}(t_2)W(t_2,t_1)R(t_1)=\sum_{m=0}^{d-1}\xi_{mm}(t_2,t_1)|m\rangle\langle m|\otimes \mathbb{I}+\sum_{m\neq n}\xi_{mn}(t_2,t_1)|m\rangle\langle n|\otimes U_{mn},
\end{align}
where $R(t)=\mathcal{T}\{e^{-i\int_{t}^{0} H'_0(t')dt'}\}=\sum_{m=0}^{d-1}|m\rangle\langle m|\otimes R_m(t)$ with $R_m(t)=\mathcal{T}\{e^{-i\int_{0}^{t}H_m(t')dt'}\}$. The ancilla jump operators become
\begin{align}\label{}
  L_l^{(I)}(t)&=R^{\dagger}(t)L_lR(t)=L_l\otimes \mathbb{I}, \nonumber \\
  J_j^{(I)}(t)&=R^{\dagger}(t)J_jR(t)=J_j\otimes R^{\dagger}_{m_j}(t)R_{n_j}(t).
\end{align}
The ancilla dephasing operator $L_l^{(I)}(t)$ remains time-independent due to $[H'_0(t),L_l]=0$. The ancilla relaxation operator $J_j^{(I)}(t)$ at most imparts a trivial phase factor to the central system if $|m_j\rangle$ and $|n_j\rangle$ are in the NAS (with $H_{m_j}(t)$ and $H_{n_j}(t)$ differing by some real constants), but implements a nontrivial unitary operation on the central system that is dependent on the jump time if $|m_j\rangle$ and $|n_j\rangle$ are not both in the NAS.

The ancilla paths can be represented in a symbolic way below, which is consistent with the schematic diagrams in Fig. {\color{blue}2} of the main text. For the ancilla starting from $|i\rangle$ and projected to $|r\rangle$, the ancilla paths without errors, with dephasing errors (e.g. $|a\rangle\langle a|$ and $|b\rangle\langle b|$) and with relaxation error (e.g. $|b\rangle\langle a|$ in NAS and $|d\rangle\langle c|\otimes R^{\dagger}_{d}(t_1)R_{c}(t_1)$ not in NAS) are denoted as
\begin{align}\label{}
   &{\rm (i)~ancilla~path~without~errors}: \nonumber \\
   &~~~|i\rangle\rightarrow|r\rangle:~~ \langle r|W^{(I)}(t,0)|i\rangle\propto U_{ri}, \\
   &{\rm (ii)~ancilla~path~with~dephasing~errors}: \nonumber \\
   &~~~|i\rangle\rightarrow|a\rangle\rightarrow|b\rangle\rightarrow|r\rangle:~~ \langle r|W^{(I)}(t,t_2)|b\rangle\langle b|W^{(I)}(t,t_1)|a\rangle\langle a|W^{(I)}(t_1,0)|i\rangle\propto U_{rb}U_{ba}U_{ai}=U_{ri}, \\
   &{\rm (iii)~ancilla~path~with~relaxation~error~in~NAS}: \nonumber \\
   &~~~|i\rangle\rightarrow|a\rangle\dashrightarrow |b\rangle\rightarrow|r\rangle:~~ \langle r|W^{(I)}(t,t_1)|b\rangle\langle a|W^{(I)}(t_1,0)|i\rangle\propto U_{rb}U_{ai}, \\
   &{\rm (iv)~ancilla~path~with~relaxation~error~not~in~NAS}: \nonumber \\
   &~~~|i\rangle\rightarrow|c\rangle\rightsquigarrow |d\rangle\rightarrow|r\rangle:~~ \langle r|W^{(I)}(t,t_1)R^{\dagger}_{d}(t_1)|d\rangle\langle c|R_{c}(t_1)W^{(I)}(t_1,0)|i\rangle\propto U_{rd}U_{ci}R^{\dagger}_{d}(t_1)R_{c}(t_1),
\end{align}
where the corresponding central system gates are the product of the unitaries after the paths. Note that each path diagram above actually represents a class of ancilla paths with all possible error times. With dephasing errors, the ancilla takes a different continuous path from $|i\rangle$ to $|r\rangle$, but the final unitary gate on the central system remains the same as that without errors, since it depends only on the initial and final states. With ancilla relaxation errors, the ancilla path is composed of discontinuous segments connected by the relaxation error operators. Moreover, for relaxation errors in NAS, the final central system gate is independent of the relaxation error time, but for relaxation errors not in NAS, the final central system gate depends the relaxation error time.

\subsubsection{Two-level ancilla with only dephasing errors}\label{2-level-de}
Consider a 2-level ancilla $\{|g\rangle, |e\rangle\}$ suffering dephasing error $\mathcal{D}[\sqrt{\gamma}(c_e|e\rangle\langle e|+c_g|g\rangle\langle g|)]$ with $c_{g/e}\in\mathbb{C}$, as shown in Fig. \ref{level}{\color{blue}(a)}. The PI no-jump propagator in Eq. (\ref{WeI}) is
\begin{align}\label{}
   W^{(I)}(t_2,t_1)=\xi_{gg}|g\rangle\langle g|\otimes \mathbb{I}+\xi_{ee}|e\rangle\langle e|\otimes \mathbb{I}
   +\xi_{ge}|g\rangle\langle e|\otimes U_{ge}+\xi_{eg}|e\rangle\langle g|\otimes U_{ge}^{\dagger},
\end{align}
where we omit the arguments $t_1, t_2$ of $\xi_{mn}(t_2,t_1) (m,n=g,e)$.

According to Theorem 1, the central system gate is PI of infinite-order ancilla dephasing errors. To understand this, supposes the ancilla starts in $|e\rangle$, the ideal case is that the central system ends in $|g\rangle$ and therefore undergoes a unitary gate $U_{ge}$ without suffering any dephasing errors during the gate. With dephasing errors, the ancilla may go around additional 1-site loops ($|g\rangle\rightarrow|g\rangle, |e\rangle\rightarrow|e\rangle$) or 2-site loop ($|e\rangle\rightarrow|g\rangle\rightarrow|e\rangle$) before arriving at $|e\rangle$ or $|g\rangle$. But since these loops produce the identity operation on the central system, so the central system will undergo a unitary gate $U_{ge}$ (the identity gate $\mathbb{I}$) with the ancilla projected to $|g\rangle$ ($|e\rangle$), even if the ancilla undergoes infinite numbers of additional loops (i.e. suffering infinite-order dephasing errors). The ancilla paths without errors, with first-order and infinite-order ancilla errors,  and the corresponding quantum gates on central system can be represented as
\begin{align}\label{}
   {\rm no~error}&: |e\rangle\rightarrow|g\rangle:~~ U_{ge}, \label{geI0}\\
   {\rm{1st~order~dephasing~error}}&\left\{ \begin{array}{l}
    \left| e \right\rangle  \to \left\{ \begin{array}{l}
    \left| g \right\rangle \\
    \left| e \right\rangle
    \end{array} \right\} \to \left| g \right\rangle :\;\;\,{U_{ge}},\\
    \left| e \right\rangle  \to \left\{ \begin{array}{l}
    \left| g \right\rangle \\
    \left| e \right\rangle
    \end{array} \right\} \to \left| e \right\rangle :\;\;\,\mathbb{I},
    \end{array} \right. \label{geI1}\\
    \vdots  \nonumber \\
    {n\rm{th~order~dephasing~error}}&\left\{ \begin{array}{l}
    \left| e \right\rangle  \to {\left\{ \begin{array}{l}
    \left| g \right\rangle \\
    \left| e \right\rangle
    \end{array} \right\}^{ \times n}} \to \left| g \right\rangle :\;\;\,{U_{ge}},\\
    \left| e \right\rangle  \to {\left\{ \begin{array}{l}
    \left| g \right\rangle \\
    \left| e \right\rangle
    \end{array} \right\}^{ \times n}} \to \left| e \right\rangle :\;\;\,\mathbb{I},
    \end{array} \right. \label{geIn}\\
    \vdots \nonumber \\
\end{align}
Thus the PI control can be repeated if the ancilla is projected to $|e\rangle$ until the gate succeeds. Similar conclusions can be reached if the ancilla starts from $|g\rangle$.

\subsubsection{Two-level ancilla with both dephasing and relaxation errors}\label{2-level}
Suppose that the 2-level ancilla also suffers the relaxation error $\mathcal{D}[\sqrt{\kappa_1}(|g\rangle\langle e|)]$ with $|g\rangle$ and $|e\rangle$ not forming a NAS, as shown in Fig. \ref{level}{\color{blue}(b)}. Then according to Theorem 2, the central system gate is not PI of any ancilla relaxation errors and therefore not PI of any ancilla errors. The reasons are twofold. First, the ancilla relaxation error produces a nontrivial operation on the central system,
\begin{align}\label{geII1}
    {\rm{1st~order~relaxation~error}}:|e\rangle\rightarrow|e\rangle\rightsquigarrow |g\rangle\rightarrow|g\rangle:~~ R^{\dagger}_{g}(t_1)R_{e}(t_1),
\end{align}
where one can see that the central system will undergoes the unitary operation $R^{\dagger}_g(t_1)R_e(t_1)$ depending on the jump time $t_1$, and therefore loses coherence after averaging over $t_1$. Second, the unitary gate $R^{\dagger}_g(t_1)R_e(t_1)$ produced by the ancilla relaxation path [Eq. (\ref{geII1})] typically differs from the desired gate $U_{ge}$ induced by the paths without any ancilla errors [Eq. (\ref{geI0})] or with only ancilla dephasing errors [Eq. (\ref{geI1}) and (\ref{geIn})] .

\subsubsection{Three-level ancilla}\label{3-level}
To make the central system gate PI of the first-order ancilla relaxation error, we can use a 3-level ancilla with states $\{|g\rangle, |e\rangle, |f\rangle\}$, where $\{|e\rangle,|f\rangle\}$ span a NAS, as shown in Fig. \ref{level}{\color{blue}(c)}. The ancilla suffers both relaxation errors $\mathcal{D}[\sqrt{\kappa_1}(|g\rangle\langle e|)]$, $\mathcal{D}[\sqrt{\kappa_2}(|e\rangle\langle f|)]$ and dephasing error $\mathcal{D}[\sqrt{\gamma_2}(c_{g}|g\rangle\langle g|+c_e|e\rangle\langle e|+c_f|f\rangle\langle f|)]$. The PI no-jump propagator in Eq. (\ref{WeI}) can be designed as
\begin{align}\label{}
   W^{(I)}(t_2,t_1)=\xi_{gg}|g\rangle\langle g|\otimes \mathbb{I}+\xi_{ee}|e\rangle\langle e|\otimes \mathbb{I}+\xi_{ff}|e\rangle\langle e|\otimes \mathbb{I}
   +\xi_{gf}|g\rangle\langle f|\otimes U_{gf}+\xi_{eg}|f\rangle\langle g|\otimes U_{gf}^{\dagger},
\end{align}
which means the central system gate is implemented by driving a $|g\rangle\leftrightarrow|f\rangle$ transition instead of the $|g\rangle\leftrightarrow|e\rangle$ transition for a 2-level ancilla. Suppose the ancilla starts in $|f\rangle$, then the ancilla paths for different kinds of ancilla errors and the corresponding central system gates can be represented as
\begin{align}\label{}
   {\rm no~error}&: |f\rangle\rightarrow|g\rangle:~~ U_{gf}, \\
   {n\rm{th~order~dephasing~error}}&\left\{ \begin{array}{l}
    \left| f \right\rangle  \to \left\{ \begin{array}{l}
    \left| g \right\rangle \\
    \left| f \right\rangle
    \end{array} \right\}^{\times n} \to \left| g \right\rangle :\;\;\,{U_{gf}},\\
    \left| f \right\rangle  \to \left\{ \begin{array}{l}
    \left| g \right\rangle \\
    \left| f \right\rangle
    \end{array} \right\}^{\times n} \to \left| f \right\rangle :\;\;\,\mathbb{I},
    \end{array} \right. \\
    {\rm 1st~order~relaxation~error}&: |f\rangle\rightarrow|f\rangle\dashrightarrow |e\rangle\rightarrow|e\rangle:~~ \mathbb{I},
\end{align}
One can see that up to the first-order ancilla relaxation and dephasing errors, the central system gate is $U_{gf}$ ($\mathbb{I}$) with the ancilla projected to $|g\rangle$ ($|e\rangle$ or $|f\rangle$). The ancilla paths caused by the second-order and third-order ancilla relaxation errors are
\begin{align}\label{}
    {\rm 2rd~order~relaxation~error}&\left\{ \begin{array}{l}
    |f\rangle  \to |f\rangle  \dashrightarrow |e\rangle  \to |e\rangle\rightsquigarrow |g\rangle  \to |g\rangle :R_g^\dag ({t_2}){R_e}({t_2}),\\
    |f\rangle  \to |f\rangle  \dashrightarrow |e\rangle  \to |e\rangle\rightsquigarrow |g\rangle  \to |f\rangle :U_{gf}^\dag R_g^\dag ({t_2}){R_e}({t_2}),
    \end{array} \right. \\
    {\rm 3rd~order~relaxation~error}&: |f\rangle\rightarrow|f\rangle\dashrightarrow |e\rangle\rightarrow|e\rangle\rightsquigarrow|g\rangle\rightarrow|f\rangle\dashrightarrow|e\rangle\rightarrow|e\rangle:~~ U_{gf}^\dag R_g^\dag ({t_2}){R_e}({t_2}).
\end{align}
From Theorem 2, one can see that the central system gate is PI of up to the first-order ancilla relaxation errors from $|f\rangle$ to $|f\rangle$ (or $|g\rangle$), and PI of up to the second-order ancilla relaxation errors from $|f\rangle$ to $|e\rangle$.

\subsubsection{$d$-level ancilla}\label{d-level}
Consider a $d$-level ancilla with states $\{|m\rangle\}_{m=0}^{d-1}$. The PI no-jump propagator in Eq. (\ref{WeI}) can be designed as
\begin{align}\label{Wpair}
 W^{(I)}(t_2,t_1)=\sum_{m=0}^{d-1} \xi_{mm}|m\rangle\langle m|\otimes \mathbb{I}
 +\sum_{\mu}\xi_{\mu}(|m_{\mu}\rangle\langle n_{\mu}|\otimes U_{\mu}+|n_{\mu}\rangle\langle m_{\mu}|\otimes U_{\mu}^{\dagger}),
\end{align}
where $\{|m_{\mu}\rangle,|n_{\mu}\rangle\}_{\mu=1}^{\mu_{\rm m}}$ ($\mu_{\rm m}<\lfloor d/2\rfloor$) are unoverlapping pairs of ancilla states. For any pair $\{|m_{\mu}\rangle,|n_{\mu}\rangle\}$, the central system undergoes the unitary evolution $U_{\mu}$ ($U_{\mu}^{\dagger}$) from $|n_{\mu}\rangle$ to $|m_{\mu}\rangle$ (from $|m_{\mu}\rangle$ to $|n_{\mu}\rangle$), while for the remaining unpaired ancilla states, the state of the central system remain unchanged, as shown in Fig. \ref{level}{\color{blue}(d)}. The ancilla states $\{|1\rangle,|3\rangle,\cdots,|2\mu_m-1\rangle, |2\mu_m+1\rangle, |2\mu_m+2\rangle,\cdots,|d\rangle\}$ form a NAS, and all the relaxation errors are within the NAS. Then up to infinite-order ancilla relaxation errors, the ancilla can go from the initial ancilla state $|i\rangle$ to the final state $|r\rangle$ ($\xi_{ri}=0$) with a unique sequence of relaxation operators in the NAS. For example, the only possible path from $|1\rangle$ to $|2\mu_m\rangle$ is
\begin{align}\label{}
    &{\rm (\mu_{m}-1)th~order~relaxation~error}: \nonumber \\
    &|1\rangle\rightarrow|1\rangle\dashrightarrow |3\rangle\rightarrow|3\rangle\dashrightarrow\cdots|2\mu_m-3\rangle\dashrightarrow|2\mu_m-1\rangle\rightarrow|2\mu_m\rangle:~~ U_{2\mu_m,2\mu_m-1}.
\end{align}
According to Theorem 2, the central system gate assisted by this $d$-level ancilla is PI of infinite-order ancilla relaxation errors and thus PI of all possible ancilla errors.

\section{General construction of PI control Hamiltonian and jump operators}

\subsection{PI control Hamiltonian}
The general effective Hamiltonian generating the PI no-jump propagator can be derived as
\begin{align}\label{}
  &H_{\rm eff}(t)=i\frac{\partial W(t,t_0)}{\partial t}W^{-1}(t,t_0)=i\frac{\partial W(t,t_0)}{\partial t}W(t_0,t) \nonumber \\
  &=i\left[\sum_{mp}^{d-1}\xi_{mp}(t,t_0)|m\rangle\langle p|\otimes \frac{\partial U_{mp}(t,t_0)}{\partial t}+\sum_{mp}^{d-1}\frac{\partial\xi_{mp}(t,t_0)}{\partial t}|m\rangle\langle p|\otimes U_{mp}(t,t_0)\right ]
    \left[\sum_{qn}^{d-1}\xi_{qn}(t_0,t)|q\rangle\langle n|\otimes U_{qn}(t_0,t) \right], \nonumber \\
  &=\sum_{m=0}^{d-1}|m\rangle\langle m|\otimes H_m(t)+i\sum_{mn}^{d-1}\left[\sum_p \frac{\partial \xi_{mp}(t,t_0)}{\partial t}\xi_{pn}(t_0,t)\right]
    |m\rangle\langle n|\otimes U_{mn}(t,t),
\end{align}
where the first part corresponds to $H_0$ or $H'_0(t)$ in the main text and the second part corresponds to $H_{\rm c}(t)-iH_{\rm jump}/2$ with $H_{\rm jump}=\sum_i \alpha_i K_i^{\dagger}K_i$ [In the derivation, we have used $\xi_{mn}(t_3,t_1)=\sum_{q=0}^{d-1}\xi_{mq}(t_3,t_2)\xi_{qn}(t_2,t_1)$, $\xi_{mn}(t,t)=\delta_{mn}$, $\partial U_{mp}(t,t_0)/\partial t=-i H_m (t)U_{mp}(t,t_0)$, $U_{mp}(t,t_0)U_{pn}(t_0,t)=U_{mn}(t,t)$].

So the general form of the PI control Hamiltonian $H_{\rm c}$ should be
\begin{align}\label{}
  H_{\rm c}(t)=\sum_{m,n}^{d-1}\varepsilon_{mn}(t)|m\rangle\langle n|\otimes U_{mn}(t,t)=\sum_{m,n}^{d-1}\varepsilon_{mn}(t)|m\rangle\langle n|\otimes R_m(t)U_{mn}R_n^{\dagger}(t),
\end{align}
where $\varepsilon_{mn}(t)=\varepsilon_{nm}^{*}(t)$.

\subsection{PI jump operators}
The jump Hamiltonian $H_{\rm jump}=\sum_i \alpha_i K_i^{\dagger}K_i$ can be in the PI form if each positive operator $K_i^{\dagger}K_i$ is in the PI form, i.e. $K_i^{\dagger}K_i=\sum_{m=0}^{d-1}\lambda_{im}|m\rangle\langle m|$ with $\lambda_{im}\geq0$ for any $m$. By polar decomposition, the general PI form of $K_i$ is
\begin{align}\label{}
K_i=S_i\sqrt{K_i^{\dagger}K_i}=\sum_{m=0}^{d-1}\sqrt{\lambda_{im}}S_i|m\rangle\langle m|,
\end{align}
with $S_i$ being a unitary matrix for the ancilla (Note that generally the ancilla jump operators $\{K_i\}$ do not commute with the PI control Hamiltonian $H_{\rm c}(t)$). The ancilla dephasing and relaxation jump operators in the main text are specific examples of the general jump operators $\{K_i\}$. The dephasing operator $L_l=\sum_{m=0}^{d-1} \Delta_{l}^{(m)}|m\rangle\langle m|=\sum_{m=0}^{d-1} |\Delta_{l}^{(m)}|S_j|m\rangle\langle m|$ with $S_l=\sum_{m=0}^{d-1}\frac{\Delta_{l}^{(m)}}{|\Delta_{l}^{(m)}|}|m\rangle\langle m|$, while the relaxation operator $J_j=|m_j\rangle\langle n_j|=S_j|n_j\rangle\langle n_j|$ with $S_j=|m_j\rangle\langle n_j|+|n_j\rangle\langle m_j|+\sum_{m\neq m_j,n_j}|m\rangle\langle m|$. The general ancilla jump operator $K_i$ can also be a superposition of different relaxation jump operators (e.g. $c_1|a\rangle\langle b|+c_2|c\rangle\langle d|$ with $c_1,c_2\in \mathbb{C}$) or a superposition of dephasing and relaxation operators (e.g. $c_1|a\rangle\langle a|+c_2|c\rangle\langle d|$).

To be PI of the ancilla error $K_i$, $S_i$ should have non-zero off-diagonal elements only in the NAS. To see this, consider
\begin{align}
 &\langle r|W(t,t_1)K_iW(t_1,0)|i\rangle \nonumber \\
 =&\sum_{n=0}^{d-1}\sum_{m=0}^{d-1}\sqrt{\lambda_{im}}\langle r|W(t,t_1)|n\rangle\langle n|S_i|m\rangle\langle m|W(t_1,0)|i\rangle \nonumber \\
 =&\sum_{n=0}^{d-1}\sum_{m=0}^{d-1}\sqrt{\lambda_{im}}\xi_{fn}(t,t_1)\xi_{mi}(t_1,0)\langle n|S_i|m\rangle R_r(t)U_{rn}R_{n}^{\dagger}(t_1)R_{m}(t_1)U_{mi},
\end{align}
For $\langle n|S_i|m\rangle\neq0$, when $|m\rangle$ and $|n\rangle$ are in the NAS with $H_m(t)$ and $H_{n}(t)$ differing by a constant, then $R_{n}^{\dagger}(t_1)R_{m}(t_1)$ is a trivial phase factor.
Then Theorem 2 in the main text still applies by replacing the ancilla relaxation jump operators $\{J_j\}$ with the general ancilla jump operators $\{K_i\}$.

\section{Exact expression of a specific PI no-jump propagator and control Hamiltonian}\label{exact}
In this section, we derive the exact expressions of a specific PI no-jump propagator and the corresponding control Hamiltonian. For simplicity, we take the interaction picture associated with $H'_0(t)=\sum_{m=0}^{d-1}|m\rangle\langle m|\otimes H_m(t)$ with the PI no-jump propagator
\begin{align}\label{WI}
 W^{(I)}(t_2,t_1)=R^{\dagger}(t_2)W(t_2,t_1)R(t_1)=\sum_{m,n}\xi_{mn}(t_2,t_1)|m\rangle\langle n|\otimes U_{mn},
\end{align}
where $R(t)=\mathcal{T}\{e^{-i\int_{t}^{0} H'_0(t')dt'}\}=\sum_{m=0}^{d-1}|m\rangle\langle m|\otimes R_m(t)$ with $R_m(t)=\mathcal{T}\{e^{-i\int_{0}^{t}H_m(t')dt'}\}$. The effective Hamiltonian in the interaction picture is
\begin{align}\label{}
 H^{(I)}_{\rm eff}(t)=&R^{\dagger}(t)H_{\rm eff}(t)R(t)-iR^{\dagger}(t)\frac{\partial R(t)}{\partial t}=H_{c}^{(I)}+H_0^{(I)}-H'_0(t)-{iH^{(I)}_{\rm jump}}/{2},
\end{align}
with $H_{c}^{(I)}=R^{\dagger}(t)H_{\rm c}(t)R(t)$, $H_{0}^{(I)}=R^{\dagger}(t)H_{0}(t)R(t)=H_0$ due to $[H_0,H'_0(t)]=0$, $H^{(I)}_{\rm jump}=\sum_l \kappa_l{\left(L_l^{(I)}\right)^\dag } {L_l^{(I)}}+\sum_j \gamma_j{\left(J_j^{(I)}\right)^\dag } {J_j^{(I)}}$ with $L_l^{(I)}/J_j^{(I)}=R^{\dagger}(t)\left(L_j/J_k\right)R(t)$. Setting $H'_0(t)=H_0$ below, we have $H^{(I)}_{\rm eff}(t)=H_{c}^{(I)}-{iH^{(I)}_{\rm jump}}/{2}$.

Consider the PI no-jump propagator for a $d$-level ancilla as shown in Fig. \ref{level}{\color{blue}(d)},
\begin{align}\label{Wpair}
 W^{(I)}(t_2,t_1)=\sum_{m=0}^{d-1} \xi_{mm}|m\rangle\langle m|\otimes \mathbb{I}
 +\sum_{\mu}\xi_{\mu}(|m_{\mu}\rangle\langle n_{\mu}|\otimes U_{\mu}+|n_{\mu}\rangle\langle m_{\mu}|\otimes U_{\mu}^{\dagger}),
\end{align}
and the control Hamiltonian producing the PI propagator, 
\begin{align}\label{FTHam}
 H_{\rm c}^{(I)}= \sum_{\mu}[\Omega_{\mu}(|m_{\mu}\rangle\langle n_{\mu}|\otimes U_{\mu}+|n_{\mu}\rangle\langle m_{\mu}|\otimes U_{\mu}^{\dagger}) 
 +(\delta_{m_{\mu}}|m_{\mu}\rangle\langle m_{\mu}|+\delta_{n_{\mu}}|n_{\mu}\rangle\langle n_{\mu}|)],
\end{align}
where the first part denotes the driving and the second part denotes the detunings.
With the PI control Hamiltonian in Eq. (\ref{FTHam}), the effective non-Hermitian Hamiltonian is
\begin{align}\label{}
H_{\rm eff}^{(I)}=&H_{\rm c}^{(I)}-\frac{i}{2}\left(\sum_l \kappa_l{(L_l^{(I)})^\dag } {L_l^{(I)}}+\sum_j \gamma_j{(J_j^{(I)})^\dag } {J_j^{(I)}}\right)
\nonumber \\
=&\sum_{\mu}[\Omega_{\mu}(|m_{\mu}\rangle\langle n_{\mu}|\otimes U_{\mu}+|n_{\mu}\rangle\langle m_{\mu}|\otimes U_{\mu}^{\dagger})
 +(\delta_{m_{\mu}}+i\lambda_{m_{\mu}})|m_{\mu}\rangle\langle m_{\mu}|+(\delta_{n_{\mu}}+i\lambda_{n_{\mu}})|n_{\mu}\rangle\langle n_{\mu}|] \nonumber \\
 &+\sum_{k\in\{\rm unpaired\}}i\lambda_k|k\rangle\langle k|,
\end{align}
where the non-Hermitian terms modify the diagonal elements. Since the ancilla states are grouped into unoverlapped pairs and single unpaired states, so the total no-jump propagator is the direct sum of the propagators for the subspaces of the paired states and unpaired states. The propagator for the unpaired states is trivial, so here we try to derive the no-jump propagator for the ancilla subspace spanned by a single pair of ancilla states.

The effective non-Hermitian Hamiltonian for the subspace $\{|m\rangle,|n\rangle\}$ (the subscripts omitted for simplicity) is
\begin{align}\label{}
 H_{{\rm eff},mn}^{(I)}= &\Omega_{mn}(|m\rangle\langle n|\otimes U_{mn}+|n\rangle\langle m|\otimes U_{mn}^{\dagger}) 
 +(\delta_{m}+i\lambda_{m})|m\rangle\langle m|+(\delta_{n}+i\lambda_n)|n\rangle\langle n|,
\end{align}
Define the identity operator and Pauli operators for the ancilla subspace $\{|m\rangle,|n\rangle\}$ as
\begin{subequations}
\begin{align}\label{}
  I_{mn}=&|m\rangle\langle m|+|n\rangle\langle n|,
  ~~~~\sigma^x_{mn}=|m\rangle\langle n|+|n\rangle\langle m|, \\
  \sigma^y_{mn}=&i(|m\rangle\langle n|-|n\rangle\langle m|),
  ~\sigma^z_{mn}=|n\rangle\langle n|-|m\rangle\langle m|,
\end{align}
\end{subequations}
then $H_{{\rm eff},mn}^{(I)}$ can be recast as
\begin{align}\label{Heffmn}
 H_{{\rm eff},mn}^{(I)}=\omega_{mn}^0I_{mn}+\omega_{mn}^{xy}\sigma_{mn}^x\otimes {\rm Re}(U_{mn})+\omega_{mn}^{xy}\sigma_{mn}^y\otimes {\rm Im}(U_{mn})+\omega_{mn}^z\sigma_{mn}^z\otimes\mathbb{I},
\end{align}
where $\omega_{mn}^0=[\delta_{m}+\delta_{n}+i(\lambda_{m}+\lambda_{n})]/2$, $\omega_{mn}^z=[\delta_{n}-\delta_{m}+i(\lambda_{n}-\lambda_{m})]/2$, $\omega_{mn}^{xy}=\Omega_{mn}$, and ${\rm Re}(U_{mn})/{\rm Im}(U_{mn})$ denotes the real/imaginary part of $U_{mn}$ [${\rm Re}(U_{mn})^2+{\rm Im}(U_{mn})^2=\mathbb{I}$ due to $U_{mn}U_{mn}^{\dagger}=U_{mn}^{\dagger}U_{mn}=\mathbb{I}$]. Then the no-jump propagator generated by $H_{{\rm eff},mn}^{(I)}$ is
\begin{align}\label{Wmn}
 &W^{(I)}_{mn}(t_2,t_1)=\exp\left[-iH_{{\rm eff},mn}^{(I)}(t_2-t_1)\right] \nonumber \\
 &=e^{-i\omega_{mn}^0(t_2-t_1)}\left\{\cosh[i\omega_{mn}(t_2-t_1)]-\sinh[i\omega_{mn}(t_2-t_1)]\left[n_{xy}\sigma_{mn}^x\otimes {\rm Re}(U_{mn})+n_{xy}\sigma_{mn}^y\otimes {\rm Im}(U_{mn})+n_z\sigma_{mn}^z\otimes\mathbb{I}\right]\right\} \nonumber \\
 &=\xi_{mm}|m\rangle\langle m|\otimes\mathbb{I}+\xi_{nn}|n\rangle\langle n|\otimes\mathbb{I}+\xi_{mn}|m\rangle\langle n|\otimes U_{mn}+\xi_{nm}|m\rangle\langle n|\otimes U_{mn}^{\dagger},
\end{align}
with
\begin{subequations}
\begin{align}\label{}
  &\xi_{mm}=e^{-i\omega_{mn}^0(t_2-t_1)}\{\cosh[i\omega_{mn}(t_2-t_1)]+n_z\sinh[i\omega_{mn}(t_2-t_1)\}, \\
  &\xi_{nn}=e^{-i\omega_{mn}^0(t_2-t_1)}\{\cosh[i\omega_{mn}(t_2-t_1)]-n_z\sinh[i\omega_{mn}(t_2-t_1)]\}, \\
  &\xi_{mn}=\xi_{nm}=e^{-i\omega_{mn}^0(t_2-t_1)}n_{xy}\sinh[i\omega_{mn}(t_2-t_1)],
\end{align}
\end{subequations}
where $\omega_{mn}=\sqrt{(\omega_{mn}^{xy})^2+(\omega_{mn}^z)^2}$, $n_z=\omega_{mn}^z/\omega_{mn}$ and $n_{xy}=\omega_{mn}^{xy}/\omega_{mn}$ (Note that $\omega_{mn}, n_z, n_{xy}\in\mathbb{C}$). In the derivation of Eq. (\ref{Wmn}), we have used $[\omega_{mn}^z\sigma_{mn}^z\otimes\mathbb{I}+\omega_{mn}^{xy}\sigma_{mn}^x\otimes {\rm Re}(U_{mn})+\omega_{mn}^{xy}\sigma_{mn}^y\otimes {\rm Im}(U_{mn})]^2=\omega_{mn}^2I_{mn}\otimes\mathbb{I}$. For hyperbolic function with complex numbers, we have the following useful formula
\begin{subequations}
\begin{align}\label{}
  &\cosh(ix)=\cos(x),\\
  &\sinh(ix)=i\sin(x), \\
  &\cosh(x+iy)=\cosh(x)\cos(y)+i\sinh(x)\sin(y),\\
  &\sinh(x+iy)=\sinh(x)\cos(y)+i\cosh(x)\sin(y).
\end{align}
\end{subequations}

If there are no ancilla errors ($\lambda_m=\lambda_n=0$) and no detuning ($\delta_m=\delta_n=0$), then $W^{(I)}_{mn}(t_2,t_1)$ becomes a simple unitary propagator
\begin{align}\label{Weffmn}
 &W^{(I)}_{mn}(t_2,t_1)=\cos\theta I_{mn}\otimes\mathbb{I}-i\sin\theta(|m\rangle\langle n|\otimes U_{mn}+|n\rangle\langle m|\otimes U_{mn}^{\dagger}),
\end{align}
with $\theta=\Omega_{mn}(t_2-t_1)$. When $\theta=\pi/2$ or $t_2-t_1=\pi/(2\Omega_{mn})$, $W^{(I)}_{mn}(t_2,t_1)=|m\rangle\langle n|\otimes U_{mn}+|n\rangle\langle m|\otimes U_{mn}^{\dagger}$ (with the trivial phase factor neglected), implying that with ancilla transition $|n\rangle\rightarrow|m\rangle$ ($|m\rangle\rightarrow|n\rangle$) a quantum gate $U_{mn}$ ($U_{mn}^{\dagger}$) is implemented on the central system.

\section{PI gates for both ancilla errors and central system errors}

Now we demonstrate that the PI propagator as in Eq. (\ref{WeI})
\begin{align}\label{}
 W^{(I)}(t_2,t_1)=\sum_{m=0}^{d-1} \xi_{mm}|m\rangle\langle m|\otimes \mathbb{I}
 +\sum_{m\neq n}\xi_{mn}|m\rangle\langle n|\otimes U_{mn},
\end{align}
can be PI of both ancilla errors (as in Theorem 1 and Theorem 2) and first-order central system errors.

One sufficient option to choose the unitary set $\{U_{mn}\}_{m,n=0}^{d-1}$ is
\begin{align}\label{}
  U_{mn}=\sum_{k}e^{i\phi_{mn,k}}F_kU_{mn,0} F_k^{\dagger}/r_{k},
\end{align}
First note that $U_{mn}$ is actually a block-diagonal unitary matrix as follows
\begin{align}\label{}
U_{mn}=\left[ {\begin{array}{*{20}{c}}
{{e^{i\phi_{mn,0}}U_{mn,0}}}&{}&{}&{}\\
{}&{{e^{i\phi_{mn,1}}U_{mn,0}}}&{}&{}\\
{}&{}& \ddots &{}\\
{}&{}&{}&{{e^{i\phi_{mn,k}}U_{mn,0}}}
\end{array}} \right]
\end{align}
where the different blocks represent the code subspace (with the projection operator $P_0$) and different errors subspaces ($\{P_k\}_{k=1}^{q-1}$ with $P_k=F_kP_0F_k^{\dagger}/r_k$) of the central system, correspondingly.
Then we have
\begin{align}\label{}
  U_{mn}F_k|\psi_0\rangle=\sum_{j}\frac{e^{i\phi_j}}{r_{j}}F_jP_0U_{mn,0}P_0 F_j^{\dagger}F_kP_0|\psi_0\rangle=e^{i\phi_k}F_kU_{mn,0}P_0|\psi_0\rangle
  =e^{i\phi_k}F_kU_{mn}|\psi_0\rangle.
\end{align}
where we have used that $P_0F_k^{\dagger}F_lP_0=r_{k}\delta_{kl}P_0$. Also the product of any possible sequence of the elements in $\{U_{mn}\}$ is in the same form as that of $U_{mn}$,
\begin{align}\label{}
  U_{ab}U_{cd}\cdots U_{em}=\sum_{k}e^{i(\phi_{ab,k}+\phi_{cd,k}+\cdots+\phi_{em,k})}F_k\left(U_{ab,0}U_{cd,0}\cdots U_{em,0}\right) F_k^{\dagger}/r_{k},
\end{align}
since the product of any block-diagonal matrix is still a block-diagonal matrix. So we also have
\begin{align}\label{product}
  U_{ab}U_{bc}\cdots U_{em}F_k|\psi_0\rangle=
  e^{i(\phi_{ab,k}+\phi_{cd,k}+\cdots+\phi_{em,k})}F_k U_{ab}U_{bc}\cdots U_{em}|\psi_0\rangle.
\end{align}

Suppose that there is no ancilla errors but a central system error $F_k^{(I)}(t)=e^{iH_0t}F_ke^{-iH_0t}=\sum_m e^{ic_{m,k}t}|m\rangle\langle m|\otimes F_k$ at time $t_1$ during the PI gate $[0,t]$, then the wavefunction of the central system after the gate is
\begin{align}
 &\langle l|W^{(I)}(t,t_1)F_k^{(I)}(t_1)W^{(I)}(t_1,0)|i\rangle|\psi_0\rangle \nonumber \\
 =&\sum_{a,b}\sum_{c,d}\sum_{m}\xi_{ab}(t-t_1)\xi_{cd}(t_1)e^{ic_{m,k}t_1}\langle l|a\rangle\langle b|m\rangle\langle m|c\rangle\langle d|i\rangle U_{ab}F_kU_{cd}|\psi_0\rangle,  \nonumber \\
 =&\sum_{m}\xi_{lm}(t-t_1)\xi_{mi}(t_1)e^{ic_{m,k}t_1}U_{lm}F_kU_{mi}|\psi_0\rangle,\nonumber \\
 =&\sum_{m}\xi_{lm}(t-t_1)\xi_{mi}(t_1)e^{i(c_{m,k}t_1+\phi_{lm,k})}F_kU_{lm}U_{mi}|\psi_0\rangle,\nonumber \\
 =&\left[\sum_{m}\xi_{lm}(t-t_1)\xi_{mi}(t_1)e^{i(c_{m,k}t_1+\phi_{lm,k})}\right]F_kU_{li}|\psi_0\rangle,
\end{align}
which implies that a single central system error happening during the gate is equivalent to the error happening after the gate and therefore can be corrected at the end of the gate. Obviously the gate is also PI of ancilla errors (permitted by Theorem 1). Thus the gate is PI of both ancilla errors and first-order central system errors, in particular, error-transparent to the first-order central system error.

\begin{figure}
\includegraphics[width=5.5in]{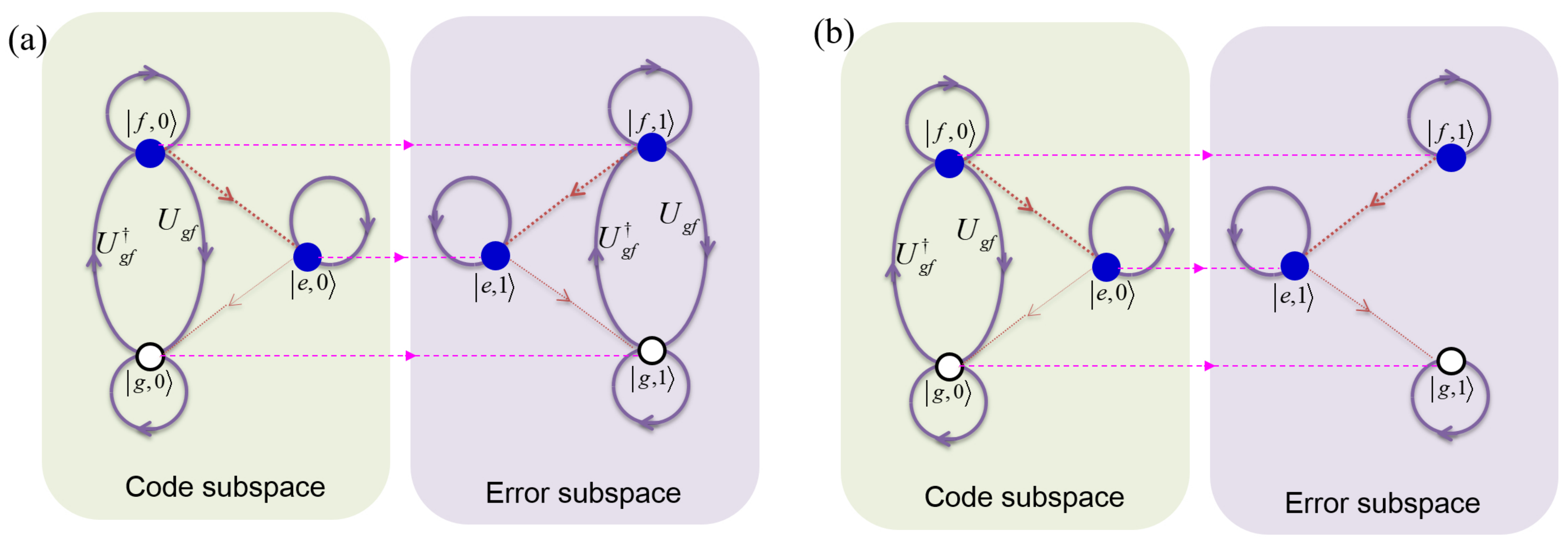}
\caption{ Schematic of PI control protocols for both ancilla errors and central system errors. (a) The driving is the same for both the code subspace and error subspace of the central system, which is error-transparent to the central system error. (b) The driving only acts on the code subspace of the central system, which is not error-transparent but still PI of the central system error. Here $|m,\mu\rangle$ denotes a generalized ancilla state, with $m=g,e,f$ denoting the three ancilla states and $\mu=0,1$ denoting the code or error subspace of the central system. The ancilla states in (out of) NAS is represented by blue solid (black hollow) circles (Strictly speaking, the circles represent the code or error subspace of the central system conditioned on a specific ancilla state), and the ancilla relaxation (central system) errors are denoted by red (magenta) dashed lines.}
\label{level2}
\end{figure}

The PI gates for both ancilla and central system errors can be understood by generalizing the PI no-jump propagator as
\begin{align}\label{WeI2}
  W^{(I)}(t_2,t_1)=\sum_{m,n=0}^{d-1}\sum_{\mu,\nu=0}^{q-1} \xi_{m\mu,n\nu}(t_2,t_1)|m,\mu\rangle\langle n,\nu|\otimes U_{m\mu,n\nu},
\end{align}
where $|m,\mu\rangle=|m\rangle\otimes|\mu\rangle$ ($m\in[0,d-1]$, $\mu\in[0,q-1]$) with $|m\rangle$ being the ancilla state and $|\mu\rangle$ denoting the different subspaces of the central system ($\mu=0$ labels the code subspace, while $\mu>0$ labels other error subspaces). Note that the above propagator in Eq. (\ref{WeI2}) can be designed to be in the same form as that in Eq. (\ref{WeI}), with the unitary set $\{U_{mk,nl}\}$ satisfying the condition $U_{m\mu,e\lambda}U_{e\lambda,n\nu}=U_{m\mu,n\nu}$ ($m,e,n\in[0,d-1]$ and $\mu,\lambda,\nu\in[0,q-1]$). Then all the conclusions (Theorem 1 and Theorem 2) of path independence for only ancilla errors can be directly applied to the cases with both ancilla and central system errors. We give a simple example [Fig. \ref{level2}] for a 3-level ancilla and a central system with a single error subspace. In Fig. \ref{level2}{\color{blue}(a)}, the driving acts on both the code subspace and error subspace so that the gate is error-transparent to first-order central system error;  In Fig. \ref{level2}{\color{blue}(b)}, the driving only acts on the code subspace of the central system, but the gate can be still PI of first-order central system error.

\section{Example: PI gates in superconducting circuits}

\begin{figure*}
\includegraphics[width=3.5in]{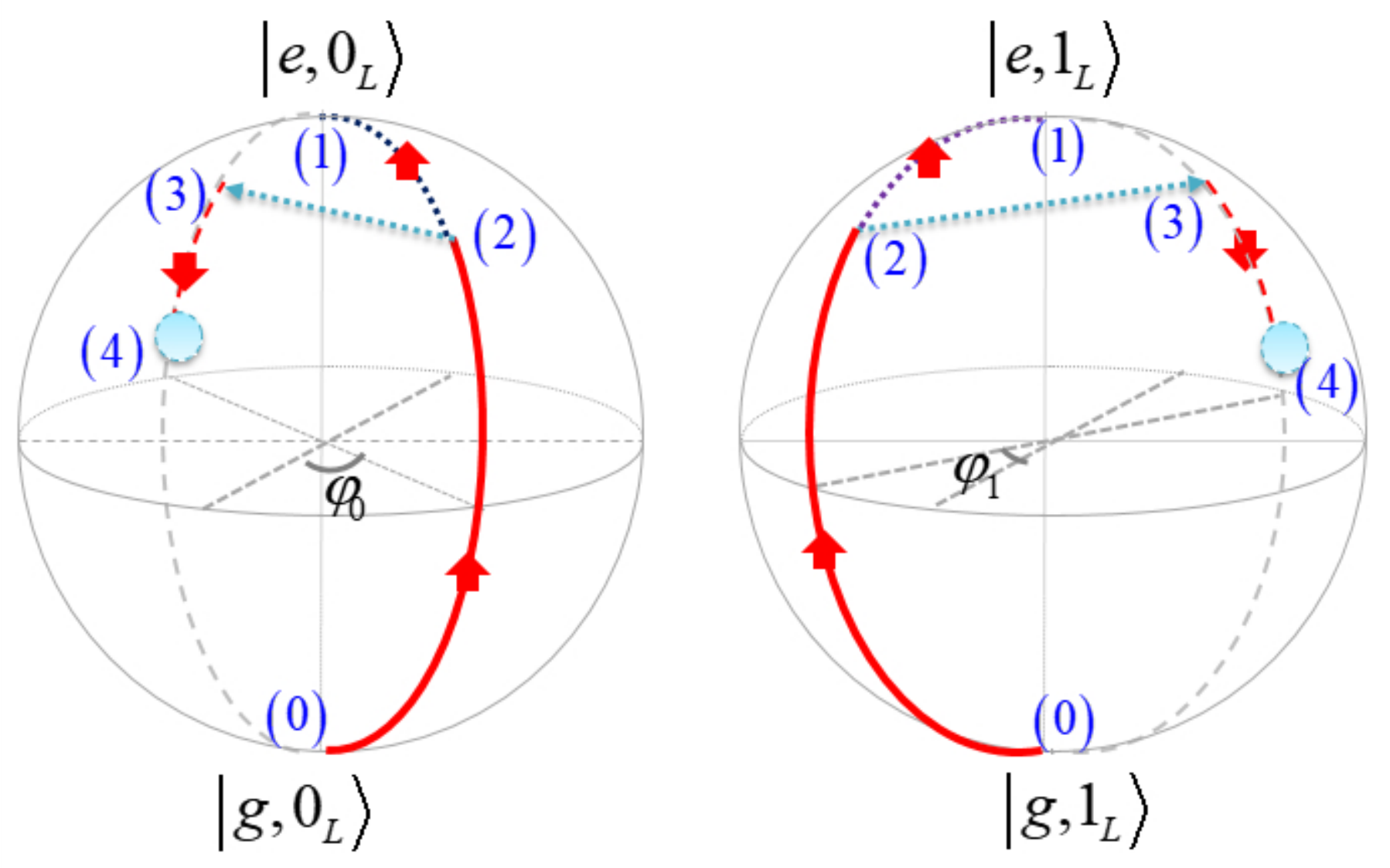}
\includegraphics[width=3.5in]{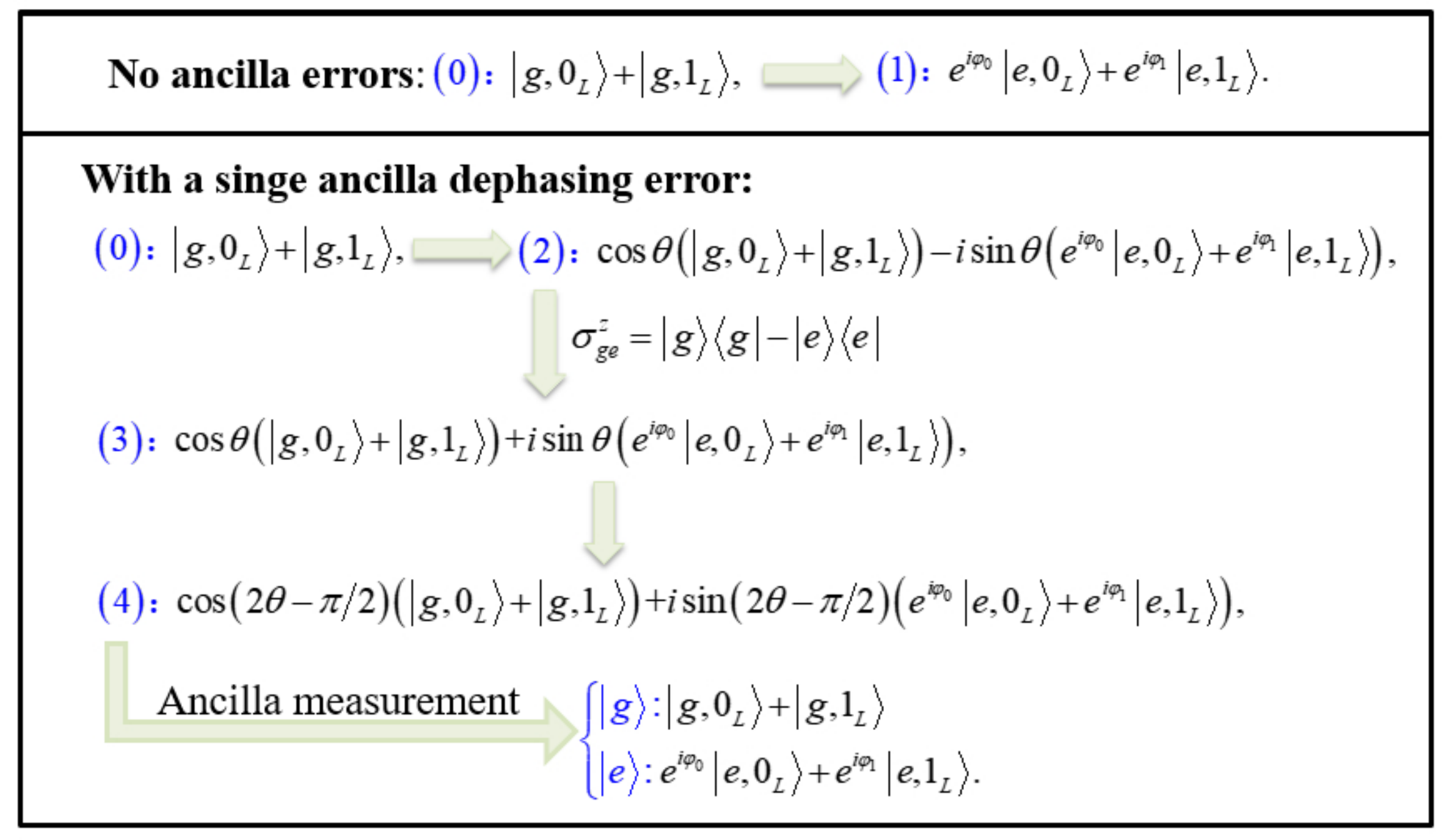}
\caption{  Schematic and state flow chart of the evolution of the 2-level transmon and a cavity mode without ancilla errors or with a single transmon dephasing error during the SNAP control. Here we take the interaction picture associated with Eq. (\ref{Hdis}) and $\{|0_L\rangle, 1_L\}$ can be any 2-dimensional subspace of the cavity mode. For the simplest binomial mode, $|0_L\rangle=(|0\rangle+|4\rangle)/\sqrt{2}$ and $|1_L\rangle=|2\rangle$. For simplicity, the states in the flow chart are not normalized.}
\label{Bloch}
\end{figure*}

In this section, we present the details of the SNAP gate in superconducting circuits as an important example of PI gates. In particular, we show that the ideal control for SNAP gate is in the PI form that can be robust against any ancilla dephasing errors. We also perform numerical simulations to demonstrate that the ideal PI control Hamiltonian for SNAP gates can be well replaced with an approximate control Hamiltonian while still significantly improving the gate performance.

The original SNAP gate consists of a set of two consecutive $\pi$-pulses (between $|g,n\rangle$ and $|e,n\rangle$ for an arbitrary set of the photon numbers $n$) with the geometric phases depending on the set of phase differences between these two $\pi$ pulses \cite{Krastanov2015,Heeres2015}. For simplicity, we may fix the phase of the set of the second pulses to be 0, so that the geometric phases for the SNAP gate are determined by the set of phases $\{\varphi_n\}$ of the first $\pi$ pulses. Moreover, we can even omit the second $\pi$ pulses, as long as we consistently keep track of the phases of later pulses relative to this choice of phase reference.

The superconducting cavity (central system) dispersively couples to a 2-level transmon device (ancilla system) with the Hamiltonian
\begin{align}\label{Hdis}
    H_{0}=\omega_{ge}|e\rangle\langle e|+\omega_{\rm c}a^{\dagger}a -\chi a^{\dagger}a|e\rangle\langle e|,
\end{align}
where $\omega_{ge}$ ($\omega_{\rm c}$) are the transmon (cavity) frequency, $a$ ($a^{\dagger}$) is the annihilation (creation) operator of the cavity mode, $\chi$ is the dispersive coupling strength, and $|e\rangle$ ($|g\rangle$) denotes the excited (ground) state of the ancilla transmon.

The SNAP gate on the cavity is
\begin{align}\label{}
    S(\vec \varphi)=\sum_{n =0}^{\infty} e^{i\varphi_n}|n\rangle\langle n|,
\end{align}
which imparts arbitrary phases $\vec \varphi=\{\varphi_n\}_{n=0}^{\infty}$ to the different Fock states of the cavity. Now we move to the interaction picture associated with $H_0$ in Eq. (\ref{Hdis}), and intend to implement a SNAP gate in this picture. The ideal control Hamiltonian is
\begin{align}\label{}
    H_{\rm c}^{(I)}=\Omega[|g\rangle\langle e|\otimes S(\vec \varphi)+|e\rangle\langle g|\otimes S(-\vec \varphi)]+\delta |e\rangle\langle e|,
\end{align}
which is a Hermitian form of the PI control Hamiltonian Eq. (\ref{Heffmn}), then from Eq. (\ref{Weffmn}) the unitary propagator induced by this Hamiltonian with $\delta=0$ can be easily found as
\begin{align}\label{Upro}
    U^{(I)}(t,0)=e^{-iH_{\rm c}^{(I)}t}=\cos\theta-i\sin\theta[|g\rangle\langle e|\otimes S(\vec \varphi)+|e\rangle\langle g|\otimes S(-\vec \varphi)],
\end{align}
where $\theta=\Omega t$. With the transmon initially in $|g\rangle$ ($|e\rangle$) and the control acting for a time $t=\pi/(2\Omega)$, the SNAP gate $S(-\vec \varphi)$ [$S(\vec \varphi)$] is perfectly implemented on the cavity with transmon ending in $|g\rangle$ ($|e\rangle$). An intuitive picture of the SNAP gate is shown in Fig. \ref{Bloch} for a logical qubit encoded in a single cavity mode. In the Schr\"{o}dinger's picture, the control Hamiltonian is
\begin{align}\label{Hc-ideal}
    H_{\rm c}^{\rm ideal}(t)=e^{-iH_0t}H_{\rm c}^{(I)}e^{iH_0t}=\Omega\sum_{n =0}^{\infty}( e^{i[(\omega_{ge}-n\chi)t+\varphi_n-\delta]}|g,n\rangle\langle e,n|+\rm H.c.).
\end{align}
With the weak-driving condition $\Omega\ll \chi$, the above control Hamiltonian can be approximated as
\begin{align}\label{Hc-approx}
    H_{\rm c}^{\rm approx}(t)=\epsilon_{ge}(t)e^{i(\omega_{ge}-\delta)t}|g\rangle\langle e|+\text{H.c.},
\end{align}
with $\epsilon_{ge}(t)=\sum_n \Omega e^{i(\varphi_n-n\chi t)}$. Such a control Hamiltonian drives the transmon alone but with multiple frequency components to distinguish the different cavity Fock states.

\begin{figure*}
\includegraphics[width=3.0in]{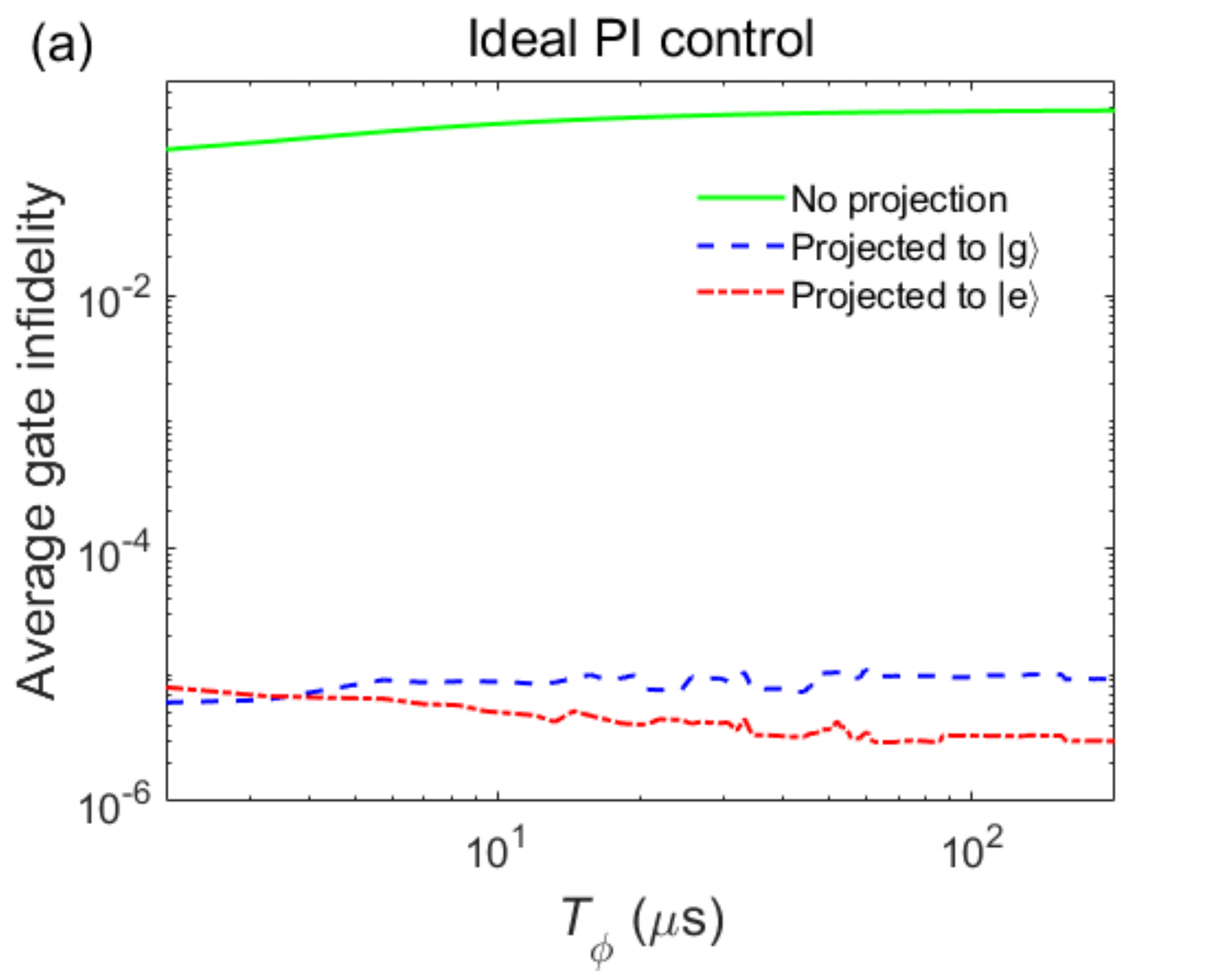}
\includegraphics[width=3.0in]{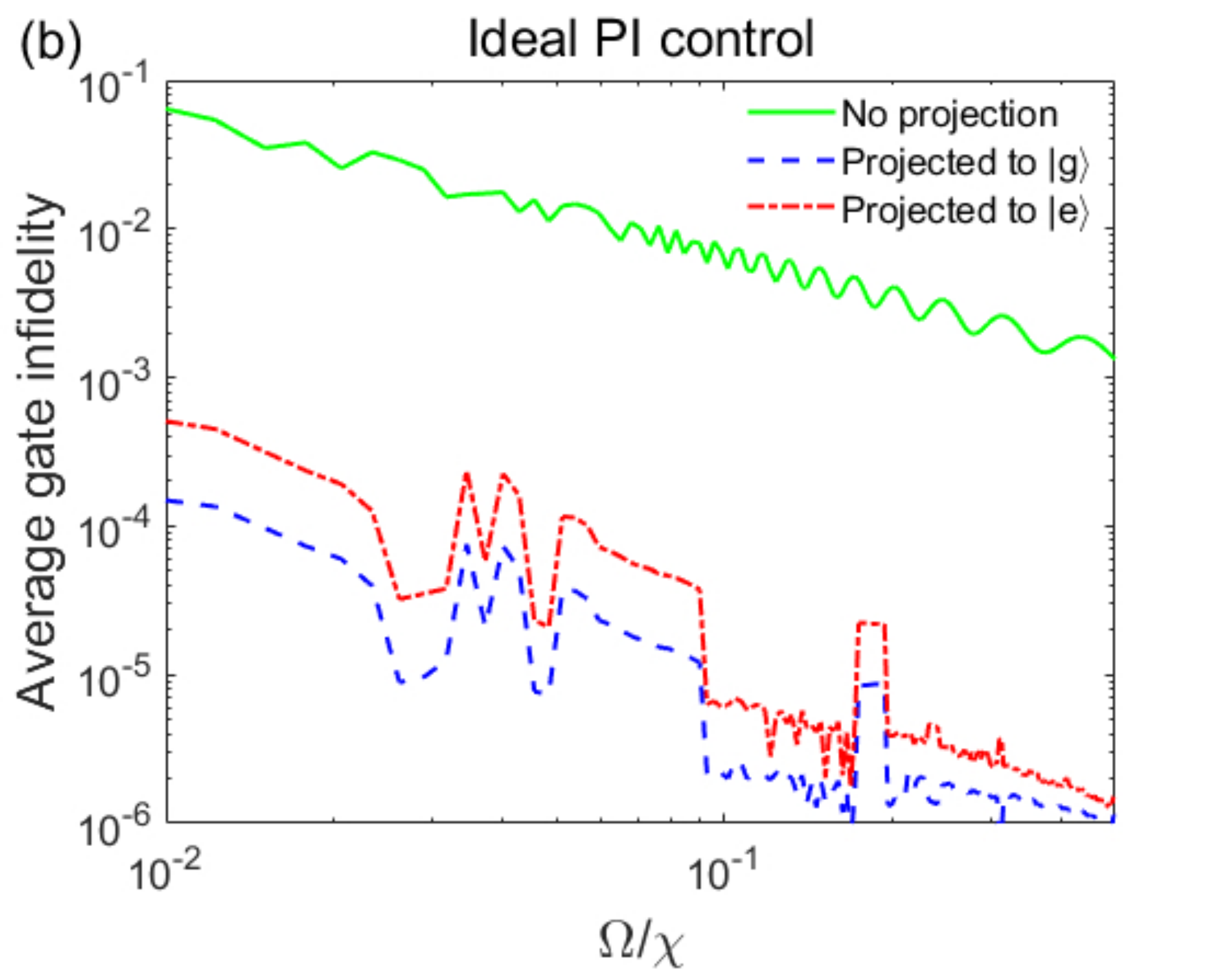}
\includegraphics[width=3.0in]{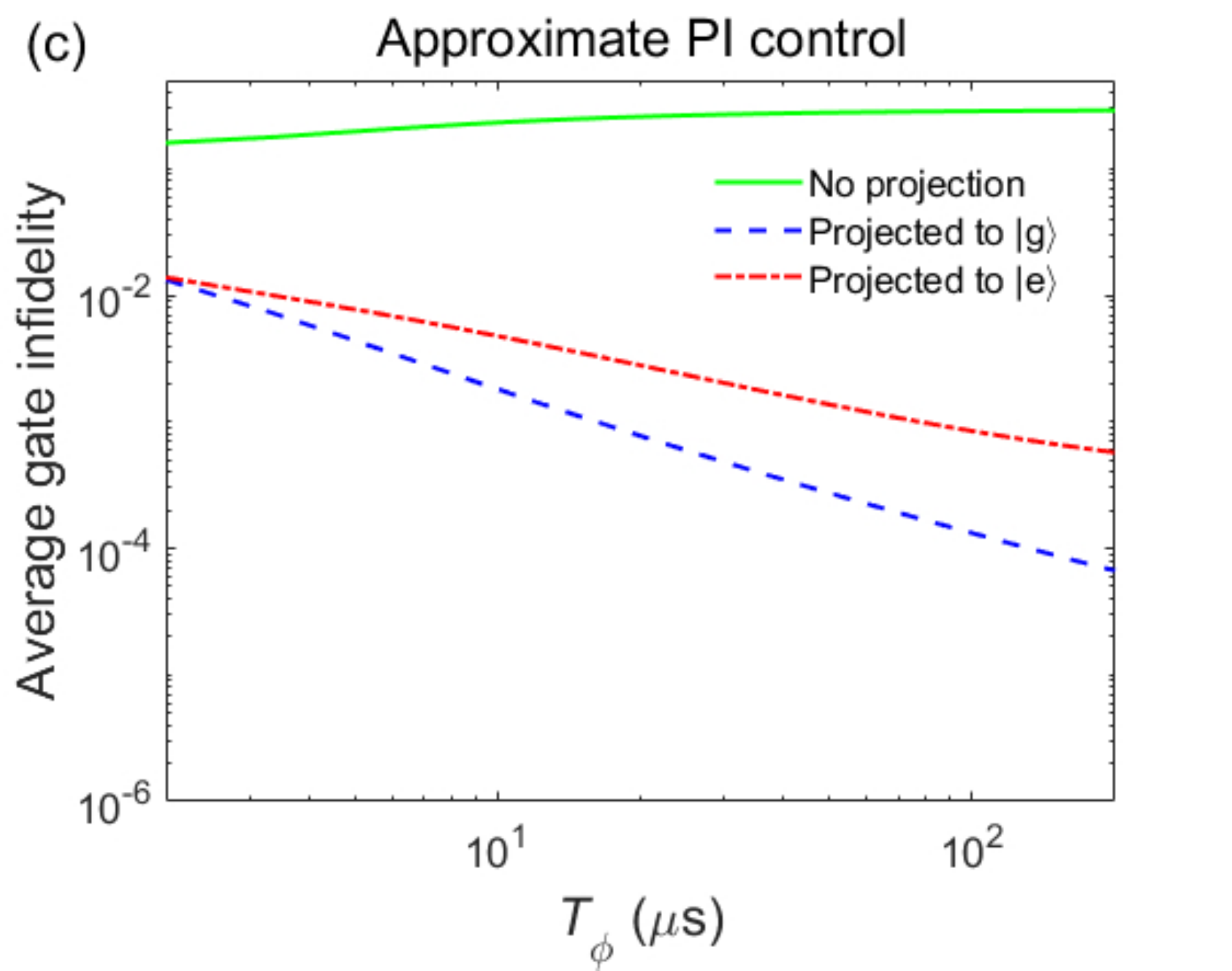}
\includegraphics[width=3.0in]{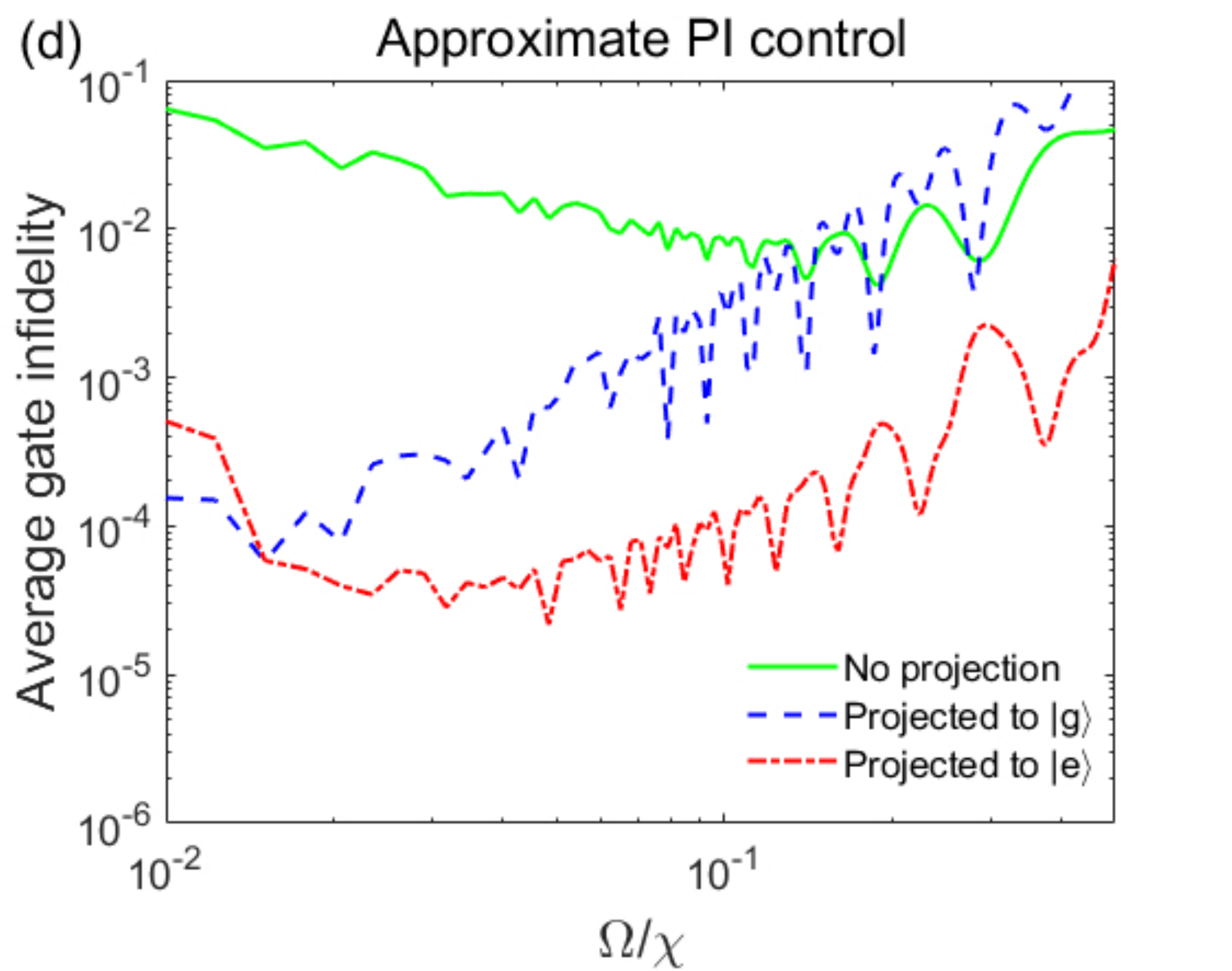}
\caption{ (a) Average gate infidelity for the ideal SNAP control [Eq. (\ref{Hc-ideal})] as a function of the transmon dephasing time $T_{\phi}$ with $\Omega=0.1$ MHz. (b) Average gate infidelity for the ideal PI SNAP control as a function of the driving amplitude $\Omega$ (in units of $\chi$) with $T_{\phi}$=70 $\mu$s. (c),(d) Similar to (a),(b) but with the approximate SNAP control [Eq. (\ref{Hc-approx})]. Here the average gate infidelity is the product of the final transmon population and the corresponding gate infidelity. With the transmon initial state $|g\rangle$, the ideal quantum operation is the SNAP (identity) gate in the interaction picture with the transmon finally projected to $|e\rangle$ ($|g\rangle$). The SNAP gate here is a $T$ gate for the simplest binomial code. The dispersive coupling is assumed to be $\chi=0.9$ MHz. }
\label{Infidelity}
\end{figure*}

Suppose that the transmon suffers only dephasing errors with the Lindbladian dissipator $\mathcal{D}[\sigma_{ge}^z/\sqrt{T_{\phi}}]$ with $\sigma_{ge}^z=|g\rangle \langle g|$ and $T_{\phi}$ being the dephasing time. With the ideal control Hamiltonian in Eq. (\ref{Hc-ideal}) and the non-Hermitian constant contributed the dephasing error [$i(\sigma_{ge}^z)^2/(2T_{\phi})=i/(2T_{\phi})$], the no-jump propagator differs from Eq. (\ref{Upro}) by a constant real factor and therefore is still in the PI form. So according to Theorem 1, the SNAP gate is PI of infinite-order transmon dephasing errors for any initial and final transmon states in the projection basis $\{|g\rangle, |e\rangle\}$ (see Sec. \ref{2-level-de}). To provide an intuitive picture, we show the state evolution of the 2-level transmon and a cavity logical qubit in Fig. \ref{Bloch}. The ideal SNAP control Hamiltonian drive the state evolves in the two Bloch subspaces $\{|g,0_L\rangle, |e,0_L\rangle\}$ and $\{|g,1_L\rangle, |e,1_L\rangle\}$ along different rotation axes but with the same speed. A single transmon dephasing error induces a simultaneous quantum jump on the two subspaces and results in incomplete rotations, but a final projective measurement of the transmon can distinguish the ideal SNAP gate or the other unitary (identity) operation depending on the measured transmon in $|e\rangle$ or $|g\rangle$. This conclusion holds even if the transmon suffers an infinite number of jumps during the control.

The robustness of the SNAP gate against the transmon dephasing error is numerically verified in Fig. \ref{Infidelity}. With the ideal SNAP control in Eq. (\ref{Hc-ideal}) and a projective measurement of transmon, the corresponding average gate infidelity as a function of $T_{\phi}$ and $\Omega/\chi$ is significantly reduced to very small value limited by the numerical errors [Fig. \ref{Infidelity}{\color{blue}(a)} and {\color{blue}(b)}]. With the approximate SNAP control in Eq. (\ref{Hc-approx}), the average gate infidelity with final transmon measurement can be still be significantly reduced compared to that without transmon measurements when $\Omega/\chi<0.1$ [Fig. \ref{Infidelity}{\color{blue}(c)} and {\color{blue}(d)}].

It should be noted that the SNAP gate with a 2-level transmon cannot be PI of the transson relaxation error $|g\rangle\langle e|$ (see Sec. \ref{2-level}). To to also PI of the dominant transmon relaxation error, we can use a 3-level transmon with $\chi$-matching condition, where the dispersive coupling strength is engineered to the same for the first-excited transmon state $|e\rangle$ and the second-excited state $|f\rangle$ (see Sec. \ref{3-level}). The SNAP gate is implemented by applying the Hamiltonian that drives the the $|g\rangle\leftrightarrow|f\rangle$ transition (enabled by the Raman drive). The experimental results in \cite{Reinhold2020} show that the SNAP gates by such a 3-level transmon with the PI design (final transmon measurement and adaptive control) achieve the reduction of logical gate error by a factor of two in the presence of naturally occurring decoherence, a sixfold suppression of the gate error with increased transmon relaxation rates and a fourfold suppression with increased transmon dephasing rates.